\documentclass[preprint]{aastex}

\def\etal{{\it et al.}~}

\begin{document}

\title{The growth of planetary embryos: orderly, runaway, or oligarchic?}

\author{R. R. Rafikov}
\affil{Princeton University Observatory, Princeton, NJ 08544}
\email{rrr@astro.princeton.edu}

\begin{abstract}
We consider the growth of a protoplanetary embryo embedded in a 
planetesimal disk. We take into account the dynamical evolution of the disk 
caused by (1) planetesimal-planetesimal interactions, which 
increase random motions
 and smooth gradients in the disk, and (2) gravitational 
scattering of planetesimals by the embryo, which tends to heat up the
disk locally and repels planetesimals away. The embryo's growth is 
self-consistently coupled to the planetesimal disk dynamics.
We demonstrate that details of the evolution depend on only 
 two dimensionless parameters incorporating all the physical characteristics 
of the problem: the ratio of the physical radius to the Hill 
radius of any solid body in the disk (which is usually a small number) 
and the number of planetesimals inside the annulus of the disk with
width equal to the planetesimal Hill radius (which is usually large). 
We derive simple scaling laws describing embryo-disk evolution for 
different sets of these parameters. The results of exploration 
in the framework of our model of 
several situations typical for protosolar nebula
can be summarized as follows: initially, the planetesimal 
disk dynamics is not affected by the presence of the embryo and the
growth of the embryo's mass proceeds very rapidly 
in the runaway regime. Later on, when the 
embryo starts being dynamically important, its accretion slows down
similar to the ``oligarchic'' growth picture. The scenario of orderly growth  
suggested by Safronov (1972) is never realized in our calculations;
scenario of runaway growth suggested by Wetherill \& Stewart (1989) is
only realized for a limited range in mass.
Slow character of the planetesimal accretion on the oligarchic stage
of the embryo's accumulation leads to a considerable increase of the  
protoplanetary 
formation timescale compared to 
that following from a simple runaway accretion  
picture valid in the homogeneous planetesimal disks.
\end{abstract}



\section{Introduction.}  
\label{sect:intro4}


It is now widely believed (e.g. Safronov 1991, Ruden 1999) 
that terrestrial planets (planets like Earth
and Venus) have formed by agglomeration of numerous rocky or icy 
bodies called planetesimals (similar to the present-day asteroids, comets,
and Kuiper Belt objects). Understanding this accretion process is central 
to understanding planet formation. 

Since the pioneering work of Safronov (1972) it has been known 
that the dynamics of the 
planetesimal disk, namely the evolution of planetesimal eccentricities 
and inclinations, is a critical factor in this process because relative 
planetesimal velocities determine to a large extent the accretion rate of
protoplanetary bodies.
The dynamical state of the disk is affected by several processes.
Gravitational scattering between planetesimals leads to the 
growth of their epicyclic energy, thereby 
 increasing relative velocities in the disk.
This interaction can  
proceed in two different regimes depending on the 
relation between the random velocity of planetesimals $v$ relative to 
local circular motion and the differential shear
across the Hill radius of two interacting bodies $r_H$:
\begin{eqnarray}
r_H=a\left(\frac{m_1+m_2}{M_c}\right)^{1/3},
\label{eq:Hill_def}
\end{eqnarray}
where $m_1$ and $m_2$ are the masses of the interacting planetesimals, 
$M_c$ is the mass of the central star and $a$ is the  
distance from the central star
at which the interaction occurs. Scattering is said to be in the 
shear-dominated
regime when $v\lesssim \Omega r_H$, and in the dispersion-dominated regime 
when $v\gtrsim \Omega r_H$ (Stewart \& Ida 2000), 
where $\Omega=\sqrt{G M_c/a^3}$ is the 
rotation frequency of a Keplerian disk. 
Scattering in the first regime is
strong for planetesimals separated by $\sim r_H$
and they can increase their random velocities very rapidly
as a result of it. In the second regime scattering is mostly weak
even at close encounters because relative velocities are large; 
in this regime it requires much more  time to change the 
kinematic properties of planetesimals. Since scattering in the 
shear-dominated case is so efficient it is likely that this
scattering would quickly heat up the disk and bring it into the 
dispersion-dominated regime. For example, a $50$-km planetesimal 
(corresponding to a mass $\approx 10^{21}$ g) at $1$ AU would enter 
the dispersion-dominated regime for interaction with other planetesimals 
when its epicyclic velocity is as small as $3$ m~s${-1}$ (corresponding
 eccentricity is only $10^{-4}$).
Thus, it is plausible that at the evolutionary stage 
when the first massive bodies start appearing 
in the system, planetesimals interact with each other in the 
dispersion-dominated regime --- an assumption which we 
will be using throughout this paper. 

Later in the course of the nebular evolution, 
 another dynamical excitation mechanism emerges:
gravitational interaction with the newly born massive bodies. We 
call these bodies embryos (following Safronov 1972)
since they are the precursors of the protoplanets; we will
require that their masses be greater than the planetesimal masses.
When these embryos become massive enough they excite
planetesimal velocities in their vicinity and also
change the spatial distribution of planetesimals around
them. The importance of these effects was first emphasized  by 
Ida \& Makino (1993), who used  N-body simulations to 
study planetesimal disks with embedded embryos. 


Depending on the dynamical state of the 
planetesimal disk, protoplanetary growth can proceed in several 
different regimes. Safronov (1972) argued that emerging embryos 
heat the planetesimal disk up to the point where the random velocities 
of constituent planetesimals are of the order of the 
escape velocity from the most 
massive embryos. It is not clear, however, whether the embryo has 
enough time to
increase planetesimal velocities to such large values given inefficiency 
of scattering in the dispersion-dominated regime.
Stewart \& Wetherill (1988) have noted the importance of
the dynamical friction in the redistribution of random energy among
planetesimals of different masses and between planetesimals and embryos. 
They argued in favor of much smaller planetesimal 
epicyclic velocities which facilitates planetary accretion.
However the spatially inhomogeneous character of the dynamical effects of the 
embryos on the disk were overlooked in their study. 

To illustrate the difference between these two scenarios 
let us estimate the embryo's accretion rate $dM_e/dt$ 
in the two-body approximation\footnote{See 
also Kokubo \& Ida (1996, 1998) for a similar discussion.} (neglecting the 
gravity of the central star):
\begin{eqnarray}
\frac{d M_e}{dt}\simeq\pi R_e^2 \Omega m N\frac{v}{v_z}
\left(1+\frac{2GM_e}{R_e v^2}\right),
\label{eq:accr_rate}
\end{eqnarray}
where $m$ is the average planetesimal mass, $M_e$ is the mass of the planet,
$R_e$ is its radius, $N$ is the surface number density of planetesimals, and 
$v_z$ is the average planetesimal velocity in the vertical 
direction (determining the disk thickness and, thus, the local volume 
density of planetesimals; it is usually of the same magnitude as $v$). 
The second term in brackets describes gravitational
focussing, which can strongly increase the collision cross-section over its
geometric value $\pi R_e^2$ when the planetesimal velocity is smaller than the
escape speed from the embryo's surface $\sqrt{2GM_e/R_e}$.
Safronov's (1972) assumption means that focussing is weak. Then one can
easily show that
\begin{eqnarray}
\frac{1}{M_e}\frac{d M_e}{dt}\propto M_e^{-1/3},
\label{eq:orderly}
\end{eqnarray} 
i.e. the embryo's growth rate slows down as its mass increases. This type
of planetary growth is known as {\it orderly} growth, because it 
implies that many embryos grow at roughly the same rate. 
On the contrary, if one
neglects the embryo's effect on $v$ and $v_z$ and assumes focussing 
to be strong
(following Wetherill \& Stewart 1988) one finds that
\begin{eqnarray}
\frac{1}{M_e}\frac{d M_e}{dt}\propto M_e^{1/3},
\label{eq:runaway}
\end{eqnarray} 
i.e. embryo's growth accelerates as its mass increases. This
corresponds to the so-called {\it runaway} accretion regime
(Wetherill \& Stewart 1989), which allows 
massive bodies grow very quickly\footnote{Formally 
the condition (\ref{eq:runaway}) means that embryo reaches infinite 
mass in finite time given unlimited supply of material 
for accretion.} in contrast 
with the orderly growth picture. 
Note that the presence of the dynamical friction is not crucial
for the onset of runaway growth (it only speeds it up) --- what is 
important is the regulation of planetesimal velocities by  
planetesimal-planetesimal scattering, rather than embryo-planetesimal 
scattering which makes $v$ and $v_z$ independent of $M_e$. 

However, it is also possible that effects of the embryo on the velocity 
dispersion are important but not so strong as orderly scenario assumes;
in this case planetesimal velocities can be increased up to several 
$\Omega R_H$, where $R_H=a(M_e/M_c)^{1/3}$ is
the Hill radius of the embryo (note that it is considerably larger than 
the Hill radius $r_H$ 
characterizing planetesimal-planetesimal scattering). 
Then formula
(\ref{eq:accr_rate}) yields equation (\ref{eq:orderly}) but with a different
proportionality constant. This regime is called {\it oligarchic} 
(Kokubo \& Ida 1998) and
is different from orderly growth because of 
 the role of the gravitational focussing
(it is very important in this regime)
and the values of planetesimal velocities (they are much smaller than the 
embryo's escape speed). As a result, more massive embryos grow slower
than the less massive ones (similar to orderly growth) but embryos
still grow faster than planetesimals in their vicinity [similar
to runaway growth, see Ida \& Makino (1993)].
The accretion in the oligarchic regime 
is considerably slower than in the runaway regime but faster than 
in the orderly regime.
For further convenience we briefly summarize major properties of 
all 3 accretion regimes:
\begin{itemize}
\item orderly --- growth of velocity dispersion in the disk is dominated 
by the embryo, gravitational focussing is weak;
\item runaway --- velocity dispersion growth is determined
solely by the planetesimal-planetesimal scattering, focussing is strong;
\item oligarchic --- velocity dispersion growth is dominated
by the embryo, focussing is strong.
\end{itemize}

The growth timescale of the embryo is a very important quantity. 
The need for the giant planets to 
accrete their huge gaseous mass onto their rocky cores 
(core-instability model, see Mizuno 1980; Stevenson 1982; 
Bodenheimer \& Pollack 1986) prior to the 
dissipation of the gaseous nebula (Hollenbach \etal 2000)
constrains the growth 
timescale in the giant planet zone  ---
between $5$ and $30$ AU in the protosolar nebula --- to be
 $\lesssim 10^6-10^7$ yr.
This interval must accommodate several distinct evolutionary
stages --- planetesimal 
formation from dust, planetesimal coagulation into protoplanetary embryos, 
and finally, catastrophic collisions between embryos to form 
rocky cores of giant planets. Thus, embryos have
only part of this time available for their growth. Orderly growth is 
characterized by very long timescales --- of order $10^8-10^9$ yr
(Safronov 1972). On the contrary,
runaway growth allows embryos to 
reach $\sim 10^{26}$ g (approximately the mass 
of the Moon) in about $10^5$ yr (Wetherill \& Stewart 1993). 
Runaway growth stops at approximately this mass because the 
embryo accretes all
the solid material within its ``feeding zone'' --- an annulus with the width 
of several $R_H$ (Lissauer 1987). 
This mass is sometimes called the {\it isolation} mass $M_{is}$
(see \S \ref{subsect:num_setup}).

The dynamical effects of the embryo on the disk, as described numerically 
by Ida \& Makino (1993) and Tanaka \& Ida (1997) and analytically 
by Rafikov (2001, 2002a, 2002b, hereafter Papers I, II, \& III 
correspondingly)
and Ohtsuki \& Tanaka (2002) are likely to change this conclusion. The results
of these studies indicate that even if the mass of the embryo is much smaller
than the isolation mass it will dominate the dynamical evolution of the
nearby region of the disk, which brings the accretion
 into the much slower oligarchic mode. In this 
case the timescale required to achieve  $M_{is}$ becomes longer than that
predicted by a simple picture of the runaway growth.

In this paper we study the coupled evolution of the planetesimal
disk and protoplanetary embryo. Our goal is to determine (1)
which conditions 
need to be fulfilled for the aforementioned regimes to be realized 
in protoplanetary nebula, (2) whether there are transitions between 
them and when do they occur, (3) which regime sets the timescale
for the planetary growth in the disk.
To answer these questions we consider a very simple model of the 
embryo-planetesimal disk system. We describe the behavior of the 
disk by representing it as a planetesimal population with a single 
nonevolving mass $m$ [similar to ``monotrophic'' model of coagulation of
Malyshkin \& Goodman (2001)]. This assumption might seem to be a gross 
oversimplification but even this toy model can still
give us a lot of insight into the details of embryo-disk dynamics. The most
important omission for the dynamics of the system which we make by employing
this assumption is the absence of dynamical friction within the  
planetesimal disk. But, as we have mentioned above, this
cannot preclude the runaway growth from happening. Also, when runaway 
or oligarchic regimes  
take place, the most massive body in the system grows faster than 
lower mass planetesimals because the growth rate increases with mass in these
accretion regimes, see equations (\ref{eq:accr_rate}) and (\ref{eq:runaway}). 
This justifies
our assumption of nonevolving planetesimal mass. We devote more 
time to the discussion of the accuracy of our 
approximations in \S \ref{sect:discussion}. 

The spatially resolved evolution of the kinematic properties of the disk 
and the growth of the embryo's mass are considered simultaneously 
and self-consistently
within the framework of our model. This distinguishes our approach from both
conventional coagulation simulations which neglect  spatial
nonuniformities of the disk properties caused by the embryo's presence
(Wetherill \& Stewart 1993; Kenyon \& Luu 1998; Inaba \etal 2001)
and purely dynamical estimates which do not allow the embryo's mass to 
vary\footnote{Probably the closest analog of our method 
is the multi-zone coagulation 
simulations of Spaute \etal (1991) and Weidenschilling \etal (1997).} 
(Ida \& Makino 1993; Paper I; Ohtsuki \& Tanaka 2002). To 
describe the coupling
between the disk dynamics and embryo's mass growth we use 
a set of equations describing both the planetesimal-planetesimal and 
the embryo-planetesimal gravitational interactions which were derived in 
Papers II \& III 
correspondingly. 
Their validity has been checked elsewhere
using direct integrations of planetesimal orbits in the 
vicinity of the embryo (Paper III). These analytical equations 
are solved numerically
in \S \ref{sect:numerical}.
In addition, to highlight the importance and better understand the 
behavior of different physical processes operating in the system 
we provide
simple scaling estimates for the evolution of the 
disk kinematic properties and the growth of the 
embryo's mass in \S \ref{sect:scaling}.
We discuss the applications and limitations of our analysis in 
\S \ref{sect:discussion}.

We will often use the model of the Minimum Mass Solar Nebula (MMSN,
Hayashi 1981) to
obtain typical numerical values of physical quantities. 
In particular, we will use the following
distribution of the surface mass density of the solids in MMSN
(Hayashi 1981):
\begin{eqnarray}
\Sigma=20~ \mbox{g~cm}^{-2}~\left(\frac{a}{1~\rm{AU}}\right)^{-3/2},
\label{eq:MMSN}
\end{eqnarray}
which is obtained assuming that solids constitute $\approx 1\%$ of
the nebular mass.


\section{Scaling arguments.}  
\label{sect:scaling}


In this section we provide simple scaling-type solutions describing
the coupled evolution of the embryo-planetesimal disk system. The reader
interested only in the numerical results of more detailed and accurate
calculations can skip this section and proceed
directly to \S \ref{sect:numerical}. 

We first derive a set of evolution equations for the embryo's mass growth 
and for the planetesimal velocity dispersions in the horizontal and vertical 
directions 
using qualitative reasoning. This will allow us to better 
understand different dynamical mechanisms operating in the disk and
predict the dependence of the system's behavior on its basic parameters.
The equations are derived for
 two different regimes of the embryo-planetesimal gravitational
interaction --- dispersion-dominated and shear-dominated, although we mostly
concentrate our attention on the former.
We will assume that planetesimal-planetesimal interactions always 
occur in the dispersion-dominated regime which almost certainly 
should be true. From now on by different velocity regimes we will imply
different regimes of the embryo-planetesimal interaction.


\subsection{Derivation of the qualitative equations.}
\label{subsect:derivation}


We start with the shear-dominated regime. In this case the 
approach velocity of a planetesimal towards the embryo 
is $\sim \Omega R_H$ and the effect of the tidal
force of the central star is quite important. As a result 
scattering is strong and each encounter with the embryo imparts 
the planetesimal with a change of its velocity 
in the horizontal direction $\Delta v\sim \Omega R_H$
(Petit \& H\'enon 1986). At the same time the 
excitation of the vertical component of the planetesimal velocity 
is weak, because the thickness of the planetesimal disk
is smaller than the embryo's Hill radius $R_H$. As a result,
scattering is almost two-dimensional and scattered planetesimals are 
restricted to the plane of the disk. Only the deviations from 
purely planar geometry can lead to the growth of inclination. On 
the basis of geometric considerations we can predict that 
$\Delta v_z\sim (v_z/\Omega R_H)\Delta v\sim v_z$. 

Planetesimals at different semimajor axes separations have different 
synodic periods (the time  between successive conjunctions 
with the embryo). In the 
shear-dominated regime the most important region of the disk is 
that within a Hill sphere of the embryo. Planetesimals in this 
region have radial separations from the embryo $\sim R_H$ and 
travel past the embryo at relative velocity $\sim \Omega R_H$.
Thus, the typical time between consecutive conjunctions 
with the embryo 
for planetesimals in
this part of the disk is
\begin{eqnarray}
T_{syn}=\frac{4\pi}{3}\Omega^{-1}\frac{a}{R_H}\sim \Omega^{-1}\frac{a}{R_H}.
\label{eq:em_syn_eeriod}
\end{eqnarray}
Then we can approximately describe the dynamical evolution of 
the planetesimal disk due to the embryo in the shear-dominated regime as 
\begin{eqnarray}
&& \frac{dv^2}{dt}\Bigg|_{sd}^{em-pl}\approx \frac{\Delta v^2}{T_{syn}}
\approx \Omega\frac{R_H}{a}(\Omega R_H)^2,
\label{eq:est_sd_v}\\
&& \frac{dv_z^2}{dt}\Bigg|_{sd}^{em-pl}\approx \frac{\Delta v_z^2}{T_{syn}}
\approx \Omega\frac{R_H}{a}v_z^2. 
\label{eq:est_sd_v_z}
\end{eqnarray}

We now turn to the dispersion-dominated regime, 
first focussing on embryo-planetesimal scattering. Here
 the two-body approximation usually proves
to be very useful. In this approximation the 
influence of the third body --- the
central star --- can be neglected and three-body scattering problem 
naturally reduces to the two-body problem which is tractable analytically.
We will carry out our qualitative arguments 
using the instantaneous scattering approximation. 
In this approximation a body with a mass $m_1$ approaching 
another body with a mass $m_2$ with relative velocity $v$ at an impact 
parameter $b$
experiences a change in velocity 
roughly equal to $\Delta v\sim G m_2/(b v)$ assuming small angle 
scattering. This approximation is valid if $b>b_0\sim G (m_1+m_2)/v^2$, since 
for $b<b_0$ large-angle scattering takes place. The upper bound on
 $b$ is set by the smaller of the disk vertical thickness 
$H\approx v_z/\Omega\gg b_0$ and the range of planetesimal epicyclic 
excursions $v/\Omega$ (usually $H$ is the one). 
Since in three-dimensional scattering 
every decade in the impact parameter $b$ between $b_0$ and $H$ provides 
roughly the same contribution to the scattering coefficients we find, 
integrating the expression for $\Delta v$ over $b$, that the
change of $v^2$ per encounter averaged over $b$ is 
\begin{eqnarray}
\langle\Delta v^2\rangle\approx \left(\frac{G m_2}{v}\right)^2
\frac{\Omega^2}{v v_z}\ln\Lambda,
~~~~~\mbox{where}~~~~\Lambda\approx H/b_0\sim \frac{v_z v^2}
{(\Omega a)^3}\frac{M_c}{m_1+m_2}.
\label{eq:av_dv2}
\end{eqnarray}
Changes of $v^2$ and $v_z^2$
are different only by constant factors because scattering
has a three-dimensional character in the dispersion-dominated regime.

Embryo-planetesimal scattering leads to changes of planetesimal velocity 
only at conjunctions with the embryo. Planetesimals which can experience close 
encounters with it occupy an annulus of the disk with width 
$\approx v/\Omega$ around the embryo's orbit. 
The typical time between the consecutive
approaches to the embryo for the planetesimals in this zone is 
$\sim T_{syn}(\Omega R_H/v)$. Then we can write  the following equation
for the velocity evolution
caused by the embryo-planetesimal encounters in the dispersion-dominated 
regime:
\begin{eqnarray}
\frac{d}{dt}\left(v^2,v_z^2\right)\Bigg|_{dd}^{em-pl}\approx 
\frac{\langle\Delta v^2\rangle}{T_{syn}}\frac{v}{\Omega R_H}
\approx \Omega \frac{R_H}{a}\frac{(\Omega R_H)^5}{v^2 v_z}\ln\Lambda_e, 
\label{eq:est_dd_em}
\end{eqnarray}
where $\Lambda_e$ is $\Lambda$ computed using $m_2=M_e\gg m=m_1$.

We next turn to planetesimal-planetesimal scattering, which always 
occurs in the dispersion-dominated regime (see \S \ref{sect:intro4}).
The volume number density of planetesimals $n$ can be expressed through the
surface number density $N$ as $n\simeq N/H\approx N\Omega/v_z$. Then we 
can write the rate of random velocity growth in the disk due to  
planetesimal-planetesimal encounters as
\begin{eqnarray}
\frac{d}{dt}\left(v^2,v_z^2\right)\Bigg|_{dd}^{pl-pl}\approx 
v n \frac{v v_z}{\Omega^2}\langle\Delta v^2\rangle
\approx \Omega N r_H^2 \frac{(\Omega r_H)^4}{v v_z}\ln\Lambda_p, 
\label{eq:est_dd_el}
\end{eqnarray}
where $m$ is the typical planetesimal mass and $\Lambda_p$ is the value of 
$\Lambda$ calculated using $m_1=m_2=m$. 
More accurate forms of the velocity growth equation (\ref{eq:est_dd_el}) 
in the dispersion-dominated
regime have been previously obtained by a number of authors 
(Stewart \& Wetherill 1988; Ida 1990; Stewart \& Ida 2000; Paper II). One
can compare them with (\ref{eq:est_dd_em}) and see that our equation 
really is a qualitative version of more accurate results. 

The embryo's accretion rate can also be approximated by the 
two-body expression
(\ref{eq:accr_rate}). We will assume that the 
physical collisions between the embryo and planetesimals occur 
with velocities at infinity 
considerably smaller than the embryo's escape velocity. This assumption,
which implies that gravitational focussing plays an important role in 
the accretion process, is usually
justified in the early stages of the embryo's growth but can 
run into problems later if planetesimal disk heats up very strongly. 
We will later check the validity of this assumption
(see \S \ref{sect:discussion}). In the 
shear-dominated regime the approach velocities of planetesimals are dominated 
by the differential shear in the disk, i.e. $v\sim \Omega R_H$. In the 
dispersion-dominated regime the approach velocities are
 dominated by the epicyclic motion 
of planetesimals. As a result we find that
\begin{eqnarray}
\dot M\Big|_{sd}\approx m N\frac{(\Omega R_H)^2}{v_z}R_e
\label{eq:shear_accr}
\end{eqnarray}
in the shear-dominated regime, and
\begin{eqnarray}
\dot M\Big|_{dd}\approx m N\frac{(\Omega R_H)^3}{v v_z}R_e
\label{eq:disp_accr}
\end{eqnarray}
in the dispersion-dominated regime. More accurate expressions for the 
accretion rate in these regimes can be found in 
Greenzweig \& Lissauer (1992) and
Dones \& Tremaine (1993).

At this point 
it is useful to switch to the following dimensionless notation
\begin{eqnarray}
&& s^2=\frac{\langle v^2\rangle}{(\Omega r_H)^2},~~~~
s_z^2=\frac{\langle v_z^2\rangle}{(\Omega r_H)^2},~~~~
{\cal M}=\frac{M_e}{m},~~~~p=\frac{R_e}{R_H},\nonumber\\
&& \tau=t/t_{syn}, ~~~~\mbox{where}~~~~
t_{syn}\equiv\frac{4\pi}{3}\Omega^{-1} \frac{a}{r_H}= T_{syn}
\frac{R_H}{r_H}=T_{syn}{\cal M}^{1/3},
\label{eq:change_of_var}
\end{eqnarray}
i.e. we normalize the planetesimal velocities by the 
shear across the planetesimal Hill
radius $r_H=a(m/M_c)^{1/3}$, time by synodic period at $1~ r_H$ and 
embryo's 
mass by the planetesimal mass  $m$. The parameter $p$ is the ratio of 
the embryo's physical radius to its Hill radius, which remains constant 
as the embryo grows at constant density. Note that in this notation
embryo-planetesimal interaction is in the shear-dominated regime when 
$s\lesssim {\cal M}^{1/3}$, and in the dispersion-dominated regime 
when $s\gtrsim {\cal M}^{1/3}$.

With these definitions we can rewrite the
equations governing the behavior of the 
 planetesimal velocity and embryo's mass growth in the 
following dimensionless form: in the shear-dominated regime
($s\lesssim {\cal M}^{1/3}$)
\begin{eqnarray}
&& \frac{d s^2}{d\tau}\approx (N a r_H)\frac{\ln\Lambda_p}{s s_z}+{\cal M},
\label{eq:s2_ev_sd}\\
&& \frac{d s_z^2}{d\tau}\approx (N a r_H)\frac{\ln\Lambda_p}{s s_z}+
s_z^2{\cal M}^{1/3},
\label{eq:sz2_ev_sd}\\
&& \frac{d {\cal M}}{d\tau}\approx p(N a r_H)\frac{{\cal M}}{s_z}.
\label{eq:mdot_sd}
\end{eqnarray}
In the dispersion-dominated regime ($s\gtrsim {\cal M}^{1/3}$)
\begin{eqnarray}
&& \frac{d}{d\tau}\left(s^2,s_z^2\right)\approx (N a r_H)
\frac{\ln\Lambda_p}{s s_z}+
{\cal M}^2\frac{\ln\Lambda_e}{s^2 s_z},
\label{eq:s2_ev_dd}\\
&& \frac{d {\cal M}}{d\tau}\approx p(N a r_H)\frac{{\cal M}^{4/3}}{s s_z}.
\label{eq:mdot_dd}
\end{eqnarray}
The different factors entering the Coulomb logarithms are
\begin{eqnarray}
\Lambda_p=s^2 s_z~~~~\mbox{and}~~~~
\Lambda_e=s^2 s_z/{\cal M}.
\label{eq:lambda}
\end{eqnarray}
In the equations of evolution of $s$ and $s_z$
(\ref{eq:s2_ev_sd}), (\ref{eq:sz2_ev_sd}), and (\ref{eq:s2_ev_dd}) the
first term on the r.h.s. describes the effect of planetesimal-planetesimal
scattering while the last one accounts for the embryo-planetesimal 
encounters. We have dropped all numerical factors entering these equations 
because our qualitative analysis cannot determine them. 
However this should not affect our general conclusions. 

We do not write down an equation of evolution for $N$ --- the surface 
density of planetesimals. Instead we simply assume it to be constant. 
This is certainly justified while the embryo's mass is small and it cannot 
perturb the spatial distribution of planetesimal orbits around it. However
as it grows bigger, the embryo starts repelling planetesimals 
(Papers I \& III) which affects their 
surface density. This effect is not included in our approach. 
It is often argued
that the conservation of the Jacobi constant of planetesimals in the course
of their scattering by the embryo can maintain their instantaneous 
surface density (which matters for the accretion) at more or 
less constant level; unfortunately it will turn out 
(see \S \ref{sect:numerical}) that this assumption is
not very good for very large embryo masses and more accurate treatment 
of evolution of $N$ is required. However, this disadvantage of
the assumption $N=const$ does not affect our major conclusions.

More accurate analysis shows that as well as the heating terms 
the equations for the planetesimal velocity evolution
should also contain transport terms. These appear because mutual 
gravitational scattering of planetesimals acts as a source of effective 
viscosity  in the disk (Papers I \& II; Ohtsuki \& Tanaka 2002) which 
tends to smooth out any inhomogeneities of planetesimal surface density 
and velocity dispersions. Equations (\ref{eq:s2_ev_sd})-(\ref{eq:s2_ev_dd})
do not take this effect into account but this does not degrade the
accuracy of our conclusions: it follows 
from our analysis of planetesimal-planetesimal scattering 
in Paper II that the transport (or viscous) terms
lead to the effects of the same magnitude as the 
planetesimal heating terms. We will 
discuss this issue in more detail in \S \ref{sect:discussion}.

The system (\ref{eq:s2_ev_sd})-(\ref{eq:lambda})
describes the dynamics of the disk only in the 
immediate vicinity of the embryo --- within $\sim R_H$ from the 
embryo's orbit in the shear-dominated regime and within 
$\sim s R_H/{\cal M}^{1/3}$ in the dispersion-dominated regime. We will call 
this region the {\it ``heated''  zone}. 
Only the dynamics of this region matters for the embryo's growth 
because embryo can accrete material only
from this part of the disk. Outside the heated region, the embryo's 
influence is negligible and disk evolution proceeds 
almost exclusively under the action 
of planetesimal-planetesimal scattering. Evolution of the distant 
parts of the disk is described by equation (\ref{eq:s2_ev_dd}) 
with the last term in its r.h.s. dropped and is rather simple.

One can easily see that the evolution equations incorporate only two free
parameters characterizing disk properties: $N a r_H$  and $p$.  What 
is more important, the more accurate 
equations derived previously in Papers II \& III
also exhibit this property (see \S \ref{sect:numerical} and 
Appendix \ref{sect:exact}). This implies that various evolutionary scenarios 
describing the growth of the embryo can be classified completely by
 these two parameters. After such classification is performed the 
general features of the evolution can be predicted 
once the values 
of the parameters $N a r_H$ and $p$ are known. 
Typical values of these parameters in 
MMSN are discussed later in \S \ref{sect:numerical}.

Our subsequent analysis of equations (\ref{eq:s2_ev_sd})-(\ref{eq:mdot_dd})
will be essentially asymptotic: we will separate the embryo-disk evolution
into a sequence of regimes in which only some terms in the evolution equations
are important. Evolution in the transition zones between separate asymptotic 
regimes will be neglected. The unimportance of the transient stages 
typically requires the changes of various quantities in asymptotic regimes
to be much larger than in the transition zones. We will see that this is 
usually true only when extreme values of parameters $N a r_H$ and $p$ are
chosen. However, even for a realistic choice of these parameters
typical for MMSN our analysis will still give us important insight into 
the coupled evolution of the embryo-disk system.


\subsection{Separation of the velocity evolution regimes.}
\label{subsect:regime_separation}


Planetesimal-planetesimal scattering dominates the evolution of the 
planetesimal velocity 
 when the first term in the r.h.s. of the corresponding equation
exceeds the second term representing the effect of the 
embryo-planetesimal scattering. The shear-dominated regime corresponds 
to ${\cal M}>s^3$ in $s-{\cal M}$ coordinates. One can see 
from (\ref{eq:s2_ev_sd}) that in this 
regime mutual planetesimal encounters dominate 
the growth of $s$ and $s_z$ when 
\begin{eqnarray}
{\cal M}<{\cal M}_{sd}(s)\approx \ln\Lambda_p\frac{N a r_H}{s^2}.
\label{eq:bound_sd}
\end{eqnarray}
In writing down this condition we have assumed that $s\sim s_z$ 
which is always
the case for three-dimensional scattering characterizing 
planetesimal-planetesimal encounters. Note that when $s\sim 1$, i.e. 
when even 
planetesimal-planetesimal scattering  is close to the shear-dominated regime,
the mass at which the embryo begins to dominate the disk dynamically
is $M_e\sim m(N a r_H)$ [$\ln\Lambda_p\sim 1$ when $s,s_z\sim 1$, 
see (\ref{eq:lambda})]. The same estimate of $M_e$ was obtained in Paper I
where both mutual planetesimal and embryo-planetesimal encounters 
were treated in the shear-dominated regime. 

The dispersion-dominated regime 
occurs when ${\cal M}<s^3$. In this case the 
disk dominates its own dynamical evolution only when [see (\ref{eq:s2_ev_dd})]
\begin{eqnarray}
{\cal M}<{\cal M}_{dd}(s)\approx 
\left(\frac{\ln\Lambda_p}{\ln\Lambda_e}\right)^{1/2}
(N a r_H)^{1/2}s^{1/2}.
\label{eq:bound_dd}
\end{eqnarray}
A similar condition (but without $\ln\Lambda_e$) was previously
derived by Ohtsuki \& Tanaka (2002).

\begin{figure}[t]
\vspace{10.5cm}
\includegraphics{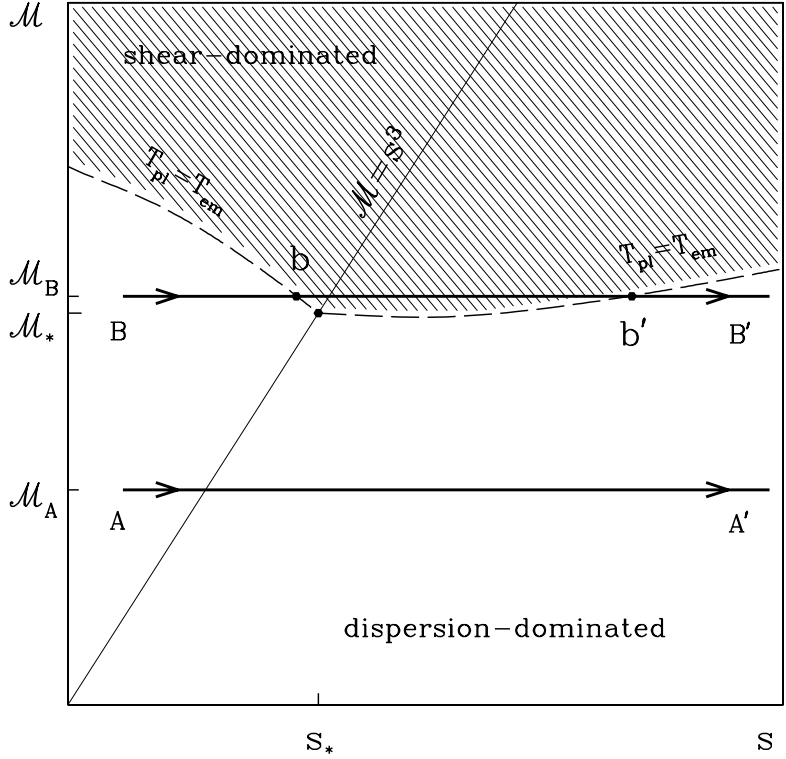}
\caption{Different regions of embryo-planetesimal interactions, 
in coordinates $s=v/(\Omega r_H)$ and ${\cal M}=M_e/m$ 
[see equation (\ref{eq:change_of_var})]. The thin solid line separates shear- 
and dispersion-dominated regimes, while the dashed curve separates regions 
where either embryo or planetesimals dominate disk heating
[see equations (\ref{eq:bound_sd}) and (\ref{eq:bound_dd}); deep in the 
dispersion-dominated regime it has a slope of $1/2$]. The embryo
controls the disk evolution in the shaded region, mutual planetesimal 
scattering dominates disk evolution in the unshaded region. Two typical 
evolutionary tracks for an embryo of fixed mass are
displayed: for ${\cal M}_A<{\cal M}_\star$ and for 
${\cal M}_B>{\cal M}_\star$.}
\label{fig:regions_no_mass}
\end{figure}

Conditions (\ref{eq:bound_sd}) and (\ref{eq:bound_dd}) separate the 
$s$-${\cal M}$ plane into regions where disk self-heating dominates
over the effect of the embryo, and vice versa. These regions are displayed
in Figure \ref{fig:regions_no_mass} and the one where embryo
controls the disk heating is shaded. The curves in the $s$-${\cal M}$ plane
given by (\ref{eq:bound_sd}) and (\ref{eq:bound_dd}) intersect with 
each other and with ${\cal M}=s^3$ at the point 
$(s_\star,{\cal M}_\star)$ where
\begin{eqnarray}
{\cal M}_\star\approx(\ln\Lambda_p)^{3/5}(N a r_H)^{3/5},~~~~
s_\star\approx(\ln\Lambda_p)^{1/5}(N a r_H)^{1/5}={\cal M}_\star^{1/3},~~~~
\Lambda_p\approx (N a r_H)^{3/5},
\label{eq:m0_s0}
\end{eqnarray}
where we have taken into account that $\ln\Lambda_e\approx 1$ when
$s,s_z\sim {\cal M}^{1/3}$, corresponding to the transition between the
shear- and dispersion-dominated regimes [see (\ref{eq:lambda})]. 

Let us assume initially that the embryo's mass does not evolve
and the disk is initially in the shear-dominated regime. Since the disk 
can only heat up, the evolutionary track of the embryo-disk system in the
$s$-${\cal M}$ plane is a straight line going to the right at
constant ${\cal M}$. It then follows from Figure \ref{fig:regions_no_mass}
that when ${\cal M}<{\cal M}_\star$, the embryo can never 
be effective at heating the
disk up and the evolution of eccentricities and inclinations of planetesimals 
takes place only as a result of 
planetesimal-planetesimal gravitational scattering
--- the effect of the embryo-planetesimal interaction is much weaker
(evolutionary track A-A$^\prime$ 
in Figure \ref{fig:regions_no_mass}). On the contrary, when 
${\cal M}>{\cal M}_\star$ (track B-B$^\prime$ 
in Figure \ref{fig:regions_no_mass}) 
embryo will start dominating the dynamical 
evolution of the heated zone when the disk is sufficiently
excited in the shear-dominated regime (at point $b$). 
As the heated zone gets more and more excited 
($s$ grows) mutual planetesimal 
scattering starts eventually to dominate the disk evolution again (at point 
$b^\prime$). This is actually not surprising because one can see
from (\ref{eq:s2_ev_dd}) that in the dispersion-dominated
regime the embryo's contribution to the disk evolution scales roughly as 
$s^{-3}$ while that of the mutual planetesimal encounters behaves as
$s^{-2}$ at a fixed ${\cal M}$. Thus, as $s$ increases planetesimal 
encounters finally take over the control over the disk evolution in 
the heated region from the embryo. It would however take a very long time 
for the system to reach the transition point $b^\prime$ because
usually $s_{b^\prime}\gg 1$ and disk 
random motion evolution is very slow in the dispersion-dominated regime.
We will not discuss any other details of the disk dynamical evolution with
fixed ${\cal M}$  because realistic embryos have $p\neq 0$ and their 
masses have to increase.


\subsection{Separation of the mass and velocity evolution regimes.}
\label{subsect:regime_separation_all}


Whenever the mass of the embryo is allowed to vary one has to study
the coupled embryo-disk evolution using the full set of
equations (\ref{eq:s2_ev_sd})-(\ref{eq:mdot_sd}) or 
(\ref{eq:s2_ev_dd})-(\ref{eq:mdot_dd}). We will concentrate here on 
exploring the dispersion-dominated regime (we briefly touch on 
the shear-dominated regime in the end of this section). 

Let us estimate different timescales associated with the evolution in the
dispersion-dominated regime. Since $s_z\sim s$, the timescale
of disk evolution caused by planetesimal-planetesimal scattering is 
[first term on the r.h.s. of (\ref{eq:s2_ev_dd})]
\begin{eqnarray}
T_{pl}=\left(\frac{1}{s}\frac{ds}{d\tau}\Big|_{pl}\right)^{-1}
\approx \frac{s^4}{\ln\Lambda_p}\frac{1}{N a r_H}.
\label{eq:pl-pl-t} 
\end{eqnarray}
The timescale characterizing the effects of the embryo is given by 
[last term on the r.h.s. of (\ref{eq:s2_ev_dd})]
\begin{eqnarray}
T_{em}
=\left(\frac{1}{s}\frac{ds}{d\tau}\Big|_{em}\right)^{-1}
\approx \frac{s^5}{\ln\Lambda_e}\frac{1}{{\cal M}^2}. 
\label{eq:em-pl-t} 
\end{eqnarray}
Finally the characteristic time of the embryo's mass growth is
\begin{eqnarray}
T_{M}=\left(\frac{1}{{\cal M}}\frac{d{\cal M}}{d\tau}\right)^{-1}
\approx \frac{s^2}{{\cal M}^{1/3}}\frac{1}{p(N a r_H)}. 
\label{eq:mass-t} 
\end{eqnarray}
By requiring $T_{pl}=T_{em}$ we retrieve the separation condition 
(\ref{eq:bound_dd}) again. 

\begin{figure}[t]
\vspace{10.5cm}
\includegraphics{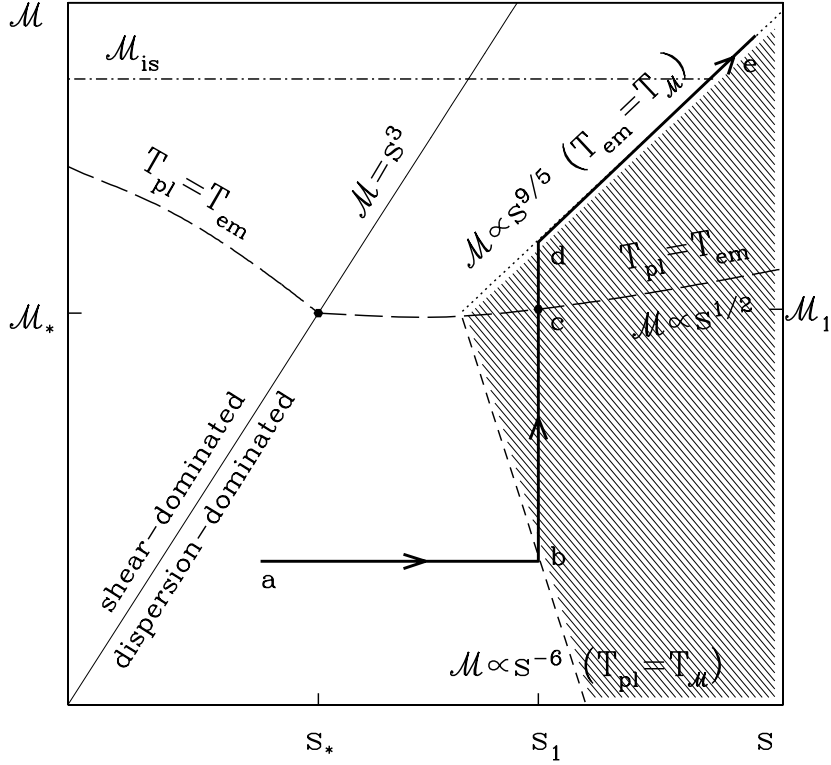}
\caption{Separation of different regions in the $s$-${\cal M}$ space
[see equation (\ref{eq:change_of_var})]
in the case of coupled evolution of disk velocity dispersion and embryo's 
mass. The dotted line displays the asymptotic 
relation (\ref{eq:m-em-separation})
and the short-dashed line shows relation (\ref{eq:m-pl-separation}). 
The dot-dashed line represents the restriction put by the existence of
the isolation mass [see (\ref{eq:isol_mass}) in \S \ref{subsect:num_setup}]. 
The embryo's
mass evolves faster than the disk velocity dispersion in the shaded region
(and vice versa in the unshaded region).
The thick solid curve shows typical evolutionary track 
of the embryo-disk system. The long-dashed curve has the same meaning 
as in Figure \ref{fig:regions_no_mass} (its slope is $1/2$ when 
$s\gg {\cal M}^{1/3}$).}
\label{fig:regions_mass}
\end{figure}

Let us assume that at the beginning embryo has mass ${\cal M}<{\cal M}_\star$
[${\cal M}_\star$ is given by equation (\ref{eq:m0_s0})]. Then initially
disk evolution is dominated by planetesimal-planetesimal scattering. 
The disk evolves faster than $M_e$ grows only when $T_{pl}\le T_M$ or
when (see Figure \ref{fig:regions_mass})
\begin{eqnarray}
{\cal M}\le {\cal M}_{M-pl}(s)\approx\left(\frac{\ln\Lambda_p}{p}
\right)^3\frac{1}{s^6}.
\label{eq:m-pl-separation}
\end{eqnarray} 
When ${\cal M}$ grows bigger than ${\cal M}_{dd}$ 
given by (\ref{eq:bound_dd}), the embryo starts dominating
the velocity evolution of the disk in the heated zone. This evolution
proceeds faster than the embryo's mass grows when $T_{em}\le T_M$ or
\begin{eqnarray}
{\cal M}\ge {\cal M}_{M-em}(s)\approx\left[\frac{p(N a r_H)}{\ln\Lambda_e}
\right]^{3/5}s^{9/5}.
\label{eq:m-em-separation}
\end{eqnarray} 
The curves in the $s$-${\cal M}$ plane given by equations 
(\ref{eq:m-pl-separation}) and (\ref{eq:m-em-separation}) intersect 
with each other and with (\ref{eq:bound_dd}) in the
dispersion-dominated regime (when $s\ge {\cal M}^{1/3}$) if $p>p_0$,
where 
\begin{eqnarray}
p_0\approx \frac{(\ln\Lambda_p)^{2/5}}{(N a r_H)^{3/5}}.
\label{eq:p_0_def}
\end{eqnarray}

Thus, when $p>p_0$ there is a region in 
the dispersion-dominated regime bounded by curves 
(\ref{eq:m-pl-separation}) and (\ref{eq:m-em-separation}) in which the
embryo's mass grows faster than the random energy in the heated zone
(shaded region in Figure \ref{fig:regions_mass}).
Everywhere outside this region the disk heats up faster than the embryo grows.
This allows us to sketch the evolution of the system
 in the following way (thick solid curve 
$a-e$ in Figure \ref{fig:regions_mass})
starting
with some ${\cal M}_0<{\cal M}_{M-pl}$: initially the disk heats up faster
than ${\cal M}$ grows 
which produces an almost horizontal track $a-b$ in the $s$-${\cal M}$ plane 
--- see Figure \ref{fig:regions_mass}. As $s$ gets to the point $b$
where ${\cal M}_0\approx{\cal M}_{M-pl}$,  mass starts growing faster
than $s$; as a result the evolution track becomes almost vertical ($b-d$). 
At point $c$ when the condition ${\cal M}={\cal M}_{dd}$ 
[see (\ref{eq:bound_dd})] is fulfilled, the embryo takes control over
the dynamical evolution of the heated zone but its mass still grows faster 
than $s$ increases [$T_{M}<T_{em}$, see (\ref{eq:em-pl-t}) and 
(\ref{eq:mass-t})].  This continues until evolutionary track reaches 
the curve ${\cal M}={\cal M}_{M-em}$ at point $d$. At this
point timescales of the embryo's growth and disk dynamical evolution 
become almost equal and embryo-disk system evolves  stably
 along the curve $d-e$. 

This general picture is confirmed by the solution of the system 
(\ref{eq:s2_ev_dd})-(\ref{eq:mdot_dd}). When ${\cal M}<{\cal M}_{dd}$ and
planetesimal-planetesimal encounters dominate the disk heating one 
finds that (treating logarithmic factors as constants)
\begin{eqnarray}
s(\tau)\approx \left[(N a r_H)\ln\Lambda_p\right]^{1/4}\tau^{1/4},~~~~
{\cal M}(\tau)\approx {\cal M}_0\left[1-\left(\tau/\tau_{run}
\right)^{1/2}\right]^{-3},
\label{eq:sols_el}
\end{eqnarray}
where 
\begin{eqnarray}
\tau_{run}\approx\frac{\ln\Lambda_p}{{\cal M}_0^{2/3}(N a r_H) p^2}
\label{eq:tau_run}
\end{eqnarray}
is the timescale of the runaway growth. One can see from this solution that
when the embryo is not massive enough to affect the disk dynamics around it
does grow in the runaway regime which is in complete agreement with 
the scenario of Wetherill \& Stewart (1989, 1993). However, in practice
the runaway solution  (\ref{eq:sols_el}) is only valid
until ${\cal M}(\tau)\approx {\cal M}_{dd}$. This mass 
is reached when $\tau\approx \tau_{run}$ since typically 
${\cal M}_{dd}\gg {\cal M}_0$. This means that embryo starts to dominate disk
heating at (see Figure \ref{fig:regions_mass})
\begin{eqnarray}
s_1=s(\tau_{run})\approx \left(\frac{\ln\Lambda_p}
{p{\cal M}_0^{1/3}}\right)^{1/2},~~~~
{\cal M}_1={\cal M}_{dd}(s_1)\approx
\left[\frac{(N a r_H)(\ln\Lambda_p)^{3/2}}
{p^{1/2}{\cal M}_0^{1/6}\ln\Lambda_e}\right]^{1/2}.
\label{eq:s1_m1}
\end{eqnarray}

After this happens evolution proceeds along a different route. 
One can find from (\ref{eq:s2_ev_dd})-(\ref{eq:mdot_dd}) 
that for ${\cal M}\gg {\cal M}_1$ the evolutionary track is described by 
the equation
\begin{eqnarray}
{\cal M}=\left[{\cal M}_1^{5/3}+\frac{p(N a r_H)}{\ln\Lambda_e}
(s^3-s_1^3)\right]^{3/5}.
\label{eq:relat_em}
\end{eqnarray}
Asymptotically, when $s\gg s_1$ this relationship reduces to 
(\ref{eq:m-em-separation}) and one obtains that
\begin{eqnarray}
s(\tau)\approx \frac{\left[p(N a r_H)\right]^{6/7}}
{(\ln\Lambda_e)^{1/7}}\tau^{5/7},~~~~
{\cal M}(\tau)\approx \frac{\left[p(N a r_H)\right]^{15/7}}
{(\ln\Lambda_e)^{6/7}}\tau^{9/7}.
\label{eq:sols_em}
\end{eqnarray}
Note that this solution does not depend on ${\cal M}_0,{\cal M}_1,s_1$, 
etc. --- all
the memory of initial conditions is lost. For this reason we would 
expect {\it any} evolutionary track describing embryo-disk 
system to behave eventually as equation (\ref{eq:sols_em}) predicts, 
independent of the initial conditions used.

The solutions of evolution equations
 obtained in this sections directly pertain only to the case $p<p_0$. 
If parameter $p$ is such that $p>p_0$ the curves given by 
(\ref{eq:m-pl-separation}) and (\ref{eq:m-em-separation}) do not intersect
in the dispersion-dominated regime. As a result embryo-planetesimal 
system can in principle switch from this velocity regime to 
the shear-dominated
or the intermediate velocity regime. We do not explicitly study 
the case $p>p_0$ in this paper ---
it turns out to be not very different from the case $p<p_0$
for the typical MMSN parameters (see \S \ref{sect:numerical}). 
We also do not derive any results 
for the embryo-disk evolution in the shear-dominated regime. 
There are several reasons for this:
\begin{itemize}
\item One would expect that initial disk velocity dispersion and
starting mass of the embryo are correlated with each other
since both require some 
time to grow; this is likely to bring the system into the dispersion-dominated
rather than shear-dominated regime. 
\item For the parameters typical for MMSN the variation of $s$ in the 
shear-dominated regime is restricted to a rather narrow range. 
It is thus possible that the asymptotic analysis such as presented in 
\S \ref{subsect:regime_separation} and 
\S \ref{subsect:regime_separation_all} would not 
be very helpful in this case since the effects 
of the transition zones between the different asymptotic 
regimes will be quite important
(see discussion in the end of \S \ref{subsect:derivation}).
This is even more important for the case $p>p_0$.
\item It is unlikely that the embryo-disk system can spend a long time 
in the shear-dominated regime since the dynamical evolution of the disk 
is usually very rapid in this case. Thus the shear-dominated evolution is
expected to be only a transient stage of the system's history.
\item The numerical approach we are going to use 
for the description of the embryo-planetesimal interaction is not 
very accurate in the shear-dominated regime (see Paper III). 
\end{itemize}

These arguments provide the basis for our present 
neglect of the shear-dominated
and intermediate velocity stages of the embryo-disk evolution.
However it is clearly not difficult to study these regimes 
qualitatively based on equations 
(\ref{eq:s2_ev_sd})-(\ref{eq:mdot_sd}) 
in the spirit of \S \ref{subsect:regime_separation} and 
\S \ref{subsect:regime_separation_all} if the need arises.


\section{Numerical results.}  
\label{sect:numerical}


One can study the embryo-disk interaction and 
check the simple qualitative predictions of \S
\ref{sect:scaling} using more robust machinery. To do this we  
use the apparatus elaborated in Papers II
\& III for the statistical description of the planetesimal-planetesimal
and embryo-planetesimal gravitational scattering. 
There we have derived a set of equations governing self-consistently both 
the planetesimal disk dynamics and the embryo's mass growth, 
taking into account
planetesimal-planetesimal as well as embryo-planetesimal gravitational
scattering. These equations are the most accurate when the embryo-planetesimal
interaction is in the dispersion-dominated regime although we are able 
to provide an approximate treatment of the intermediate velocity regime 
as well (see Paper III). In Appendix \ref{sect:exact} we transform this set of 
equations into a dimensionless form. We then numerically solve
them and find the embryo's mass as a function of time; in addition we 
obtain time-dependent and spatially-resolved profiles of the planetesimal
surface density and  dispersions of eccentricity and inclination. 
What makes this analysis novel is that it takes into account
the effects of planetesimal disk nonuniformities induced by the 
embryo's gravity. 
The closest existing analog of our method is a multi-zone coagulation
simulation approach of Spaute \etal (1991) and Weidenschilling \etal (1997)
which achieves spatial resolution by employing a large number of 
single-zone coagulation simulations interacting through the boundaries. 
However in their published simulations the treatment of the disk 
dynamical evolution is very rough and the degree of the spatial resolution 
is not very large --- of the order of Hill radii of biggest bodies at best 
while our approach allows us to use much smaller grid sizes. 


\subsection{Numerical setup.}
\label{subsect:num_setup}


In this paper we are interested in the growth of a single isolated embryo.
We assume a planetesimal population with a single characteristic mass
and disregard the evolution of this mass due to the planetesimal coagulation.
We neglect any mechanisms which can affect planetesimal dynamics 
(gas drag, inelastic collisions, etc.) other than gravitational
scattering. The embryo's eccentricity and
 inclination are assumed to be always zero. We are
going to relax these assumptions in future work.
As the embryo's mass increases, its Hill radius $R_H$ 
(which is a characteristic length scale  of the problem) grows as well.
For this reason we compute the spatial distributions of various quantities 
 not in the physical coordinates but in Hill coordinates of the embryo which
naturally adjust to the embryo's mass growth. For the same reason 
in our numerical calculations we
characterize the dynamical state of planetesimal disk not by 
$s$ and $s_z$ defined by (\ref{eq:change_of_var})
but by $S$ and $S_z$ --- velocity dispersions
in horizontal and vertical directions normalized by the embryo's Hill
radius $R_H$ (and not by the planetesimal Hill radius $r_H$):
\begin{eqnarray} 
S^2=\frac{\langle v^2\rangle}{(\Omega R_H)^2}=
\frac{\sigma_e^2}{\mu_e^{2/3}}=\frac{s^2}{{\cal M}^{2/3}},
~~~S_z^2=\frac{\langle v_z^2\rangle}{(\Omega R_H)^2}
=\frac{\sigma_i^2}{\mu_e^{2/3}}=\frac{s_z^2}{{\cal M}^{2/3}},
\label{eq:new_vels}
\end{eqnarray}
where $\sigma_e$ and $\sigma_i$ are the (nonreduced) eccentricity and 
inclination dispersions of Paper II.
When making comparisons with \S \ref{sect:scaling} we usually switch back
to variables $s$ and $s_z$. The dimensionless variables ${\cal M}$ and
$\tau$ defined in equation (\ref{eq:change_of_var}) are used instead of the
embryo's mass and time.
The width of the integration region in Hill coordinates 
is chosen to be large enough [usually $(-40 R_H; 40 R_H)$ in coordinates 
which adjust to the embryo's mass] 
so as to minimize boundary effects. 

To provide an estimate of the random velocity dispersion $s$
in the heated zone (which is one of the functions of interest)
in our numerical solutions we have chosen the following method:  
we determine the extent of the 
region around the embryo within which the excess of 
eccentricity dispersion over its value at infinity is $>20\%$ of the maximum 
value of this excess. We define this to be the heated zone.
Then we average $(S^2+S_z^2)^{1/2}$ over this region 
and use the resultant quantity [multiplied by ${\cal M}^{2/3}$, see
(\ref{eq:new_vels})]
as a measure of planetesimal 
random velocity $s$. At this point we do not make a distinction between 
the total velocity $\sqrt{s^2+s_z^2}$ and its horizontal component $s$
because they are approximately the same at our level of accuracy. 
This is rather loose definition of $s$ but it suffices 
for our purposes.

\begin{center}
\begin{deluxetable}{ l l l l }
\tablecolumns{4}
\tablewidth{0pc}
\tablecaption{Typical values of $N a r_H$ and $p$.\tablenotemark{1}  
\label{table}}
\tablehead{
\colhead{Typical object}&
\colhead{$a$, AU}&
\colhead{$Nar_H$ ($m=10^{21}$ g)}&
\colhead{$p$}
	}
\startdata
Earth  & $1$ & 
$360$ & $3.6\times 10^{-3}$ \\
Jupiter core  & $5$ & 
$800$ & $7\times 10^{-4}$ \\
KBO  & $40$ & 
$2.3\times 10^3$ & $0.9\times 10^{-4}$\\
\enddata
\tablenotetext{1}{In computing these values $M_c=M_\odot$ and 
$\rho=3$ g cm$^{-3}$ are used.}
\end{deluxetable}
\end{center}

As we have already mentioned in \S \ref{subsect:derivation} the outcome 
of the embryo and planetesimal disk evolution depends only on the 
values of $N a r_H$ and $p$. This simplifies our treatment a lot by 
combining several different quantities (such as $m$, $a$, $N$, $M_c$,
etc.)
into just two parameters.
The parameter $N a r_H$ has the meaning of (roughly) 
the number of planetesimals
within the annulus of width $r_H$ and radius equal to the embryo's semimajor
axis $a$. This number is typically quite large. Parameter $p$ is simply the
ratio of the physical size of the body to its Hill radius. Since both radii 
depend on the mass of the body in the same way
(both are $\propto M_e^{1/3}$) $p$ is actually independent of 
$M_e$ and can be taken to characterize both the embryo and 
planetesimals. In many astrophysically relevant 
situations this parameter is small. Using MMSN parameters given by 
(\ref{eq:MMSN}) we can estimate that
\begin{eqnarray}
N a r_H =\frac{\Sigma a^2}{m^{2/3}M_c^{1/3}}\approx
360\left(\frac{a}{1~\rm{AU}}\right)^{1/2}
\left(\frac{10^{21}~\rm{g}}{m}\right)^{2/3}
\left(\frac{M_\odot}{M_c}\right)^{1/3}
\label{eq:nar_H}
\end{eqnarray}
and 
\begin{eqnarray}
p =\left(\frac{3}{4\pi}\frac{M_c}{\rho a^3}\right)^{1/3}\approx
3.6\times 10^{-3}\left(\frac{1~\rm{AU}}{a}\right)
\left(\frac{3~\rm{g~cm}^{-3}}{\rho}\right)^{1/3}
\left(\frac{M_c}{M_\odot}\right)^{1/3},
\label{eq:p}
\end{eqnarray}
where $\rho$ is the density of solid material constituting 
planetesimals as well as the embryo.
In Table 1 one can find typical values of these parameters
in different important locations of the protosolar nebula.
Note that for a given $\rho$ and $M_c$ equation (\ref{eq:p}) unambiguously
determines the semimajor axis in the MMSN corresponding to a 
particular value of $p$. Knowledge of $N a r_H$ then allows one to 
fix the planetesimal mass $m$.  

We numerically solve the evolution equations of Appendix \ref{sect:exact}
and compare the results with the predictions of \S \ref{sect:scaling}. 
We display the outcomes of calculations for 3 
representative cases: $N a r_H=400, p=0.004$
corresponding to $a=0.9$ AU and $m=7.3\times 10^{20}$ g which is 
typical for the terrestrial planet region;
$N a r_H=10^3, p=10^{-3}$ ($a=3.6$ AU and $m=1.5\times 10^{21}$ g)
which is close to the region of giant planet formation; 
$N a r_H=10^3, p=10^{-4}$
($a=36$ AU and $m=4.7\times 10^{22}$ g) which is representative of the 
distant part of the protosolar nebula where Kuiper Belt Objects (KBOs)
are assumed to form (although the typical planetesimal mass is 
almost certainly 
too big in this case). In each case the calculation was carried out
 for several values 
of initial embryo's mass ${\cal M}_0$ and velocity dispersion $s_0$
(which is usually unimportant).
Each calculation starts 
with a protoplanetary embryo in a homogeneous disk; we follow both 
the subsequent growth 
of embryo's mass and the development of disk inhomogeneities. 

We allow the disk-embryo system to evolve until ${\cal M}$ 
exceeds the restriction
put by the existence of the isolation mass $M_{is}$ (for higher masses 
our results are not accurate because planetesimal surface density would be 
considerably reduced by accretion). From the definition
of $M_{is}$ (see \S \ref{sect:intro4}) it follows that\footnote{We assume here 
that the width of the feeding zone is $2R_H$ rather than $R_H$ as has 
been assumed in Paper I.} $M_{is}\approx
2\pi a\times 2a(M_{is}/M_c)^{1/3}m N$, or in our dimensionless notation 
\begin{eqnarray}
{\cal M}_{is}\simeq (4\pi)^{3/2}(N a r_H)^{3/2}.
\label{eq:isol_mass}
\end{eqnarray}
In fact, in the dispersion-dominated regime this condition should
 be additionally multiplied by $S^{3/2}=s^{3/2}/{\cal M}^{1/2}$ because
in this velocity regime the width of the 
feeding zone is about the planetesimal
epicyclic excursion $SR_H$. Thus, (\ref{eq:isol_mass}) should
be considered a lower limit on the isolation mass which is valid in the 
shear-dominated regime only; in the dispersion-dominated regime the 
isolation mass becomes
\begin{eqnarray}
{\cal M}_{is}\simeq 4\pi(N a r_H) s.
\label{eq:isol_mass_dd}
\end{eqnarray}
Obviously the more dynamically excited planetesimal disk is (i.e. the higher
$s$ is) the more massive the embryo can grow, 
because it can accrete planetesimals
from a larger region of the nebula.

\begin{figure}[t]
\vspace{10.5cm}
\includegraphics{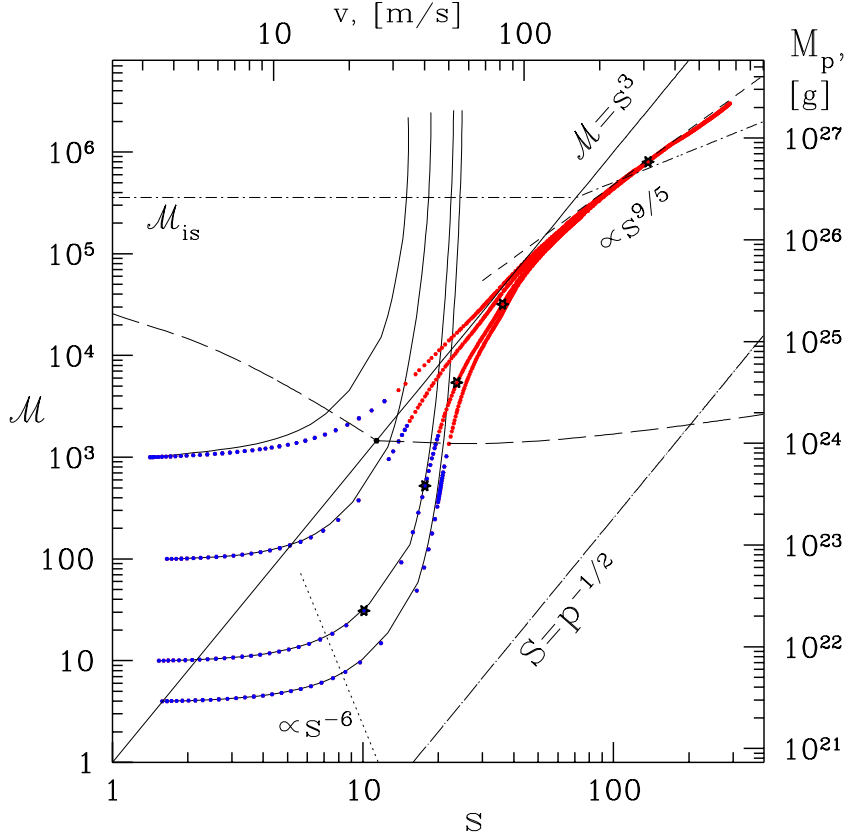}
\caption{Evolution of the embryo-disk system in $s$-${\cal M}$ coordinates
for $N a r_H=400$ and $p=0.004$ (typical for the 
Earth-forming region of MMSN).
The evolutionary tracks  
consist of two portions: where planetesimal scattering dominates the 
dynamical evolution of the disk (blue dots)
and where embryo dominates this process (red dots).
Curves separating various regions in this plot have the same meaning as
in Figure \ref{fig:regions_mass}. The thin solid curves initially
coinciding with the dotted tracks describe the runaway growth of 
``passive'' embryos.}
\label{fig:evol_2.6_2.4}
\end{figure}

\begin{figure}[t]
\vspace{10.5cm}
\includegraphics{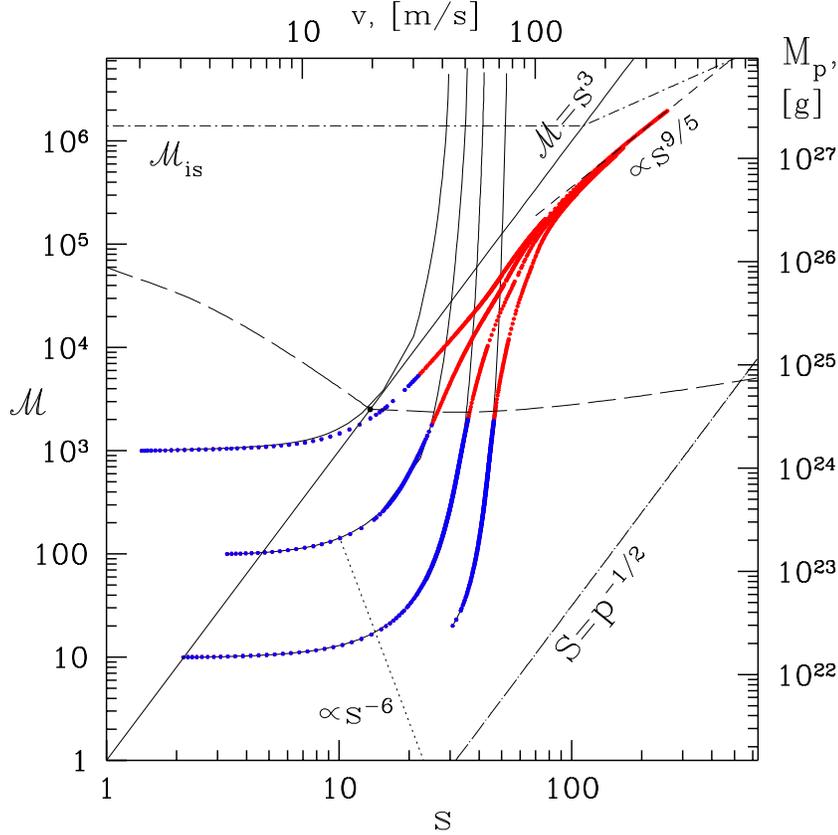}
\caption{
The same as Figure \ref{fig:evol_2.6_2.4} but 
for $N a r_H=10^3$ and $p=10^{-3}$ (typical for the Jupiter-forming 
region of MMSN).}
\label{fig:evol_3_3}
\end{figure}

\begin{figure}[t]
\vspace{10.5cm}
\includegraphics{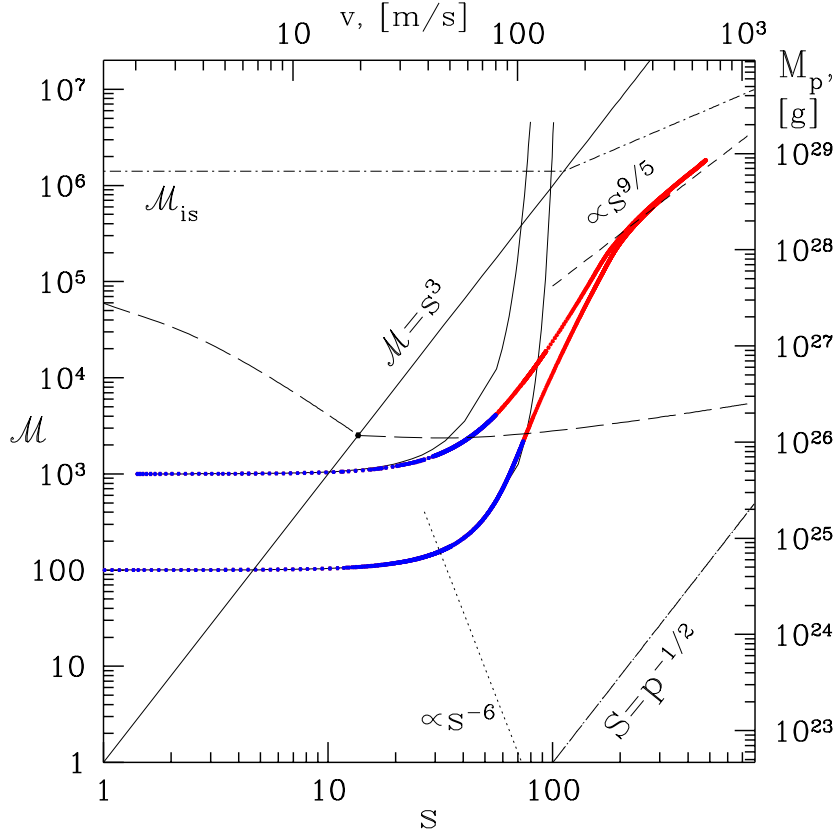}
\caption{The same as Figure \ref{fig:evol_2.6_2.4} but 
for $N a r_H=10^3$ and $p=10^{-4}$ (typical for the KBO-forming 
region of MMSN).}
\label{fig:evol_3_4}
\end{figure}

\begin{figure}[t]
\vspace{10.5cm}
\includegraphics{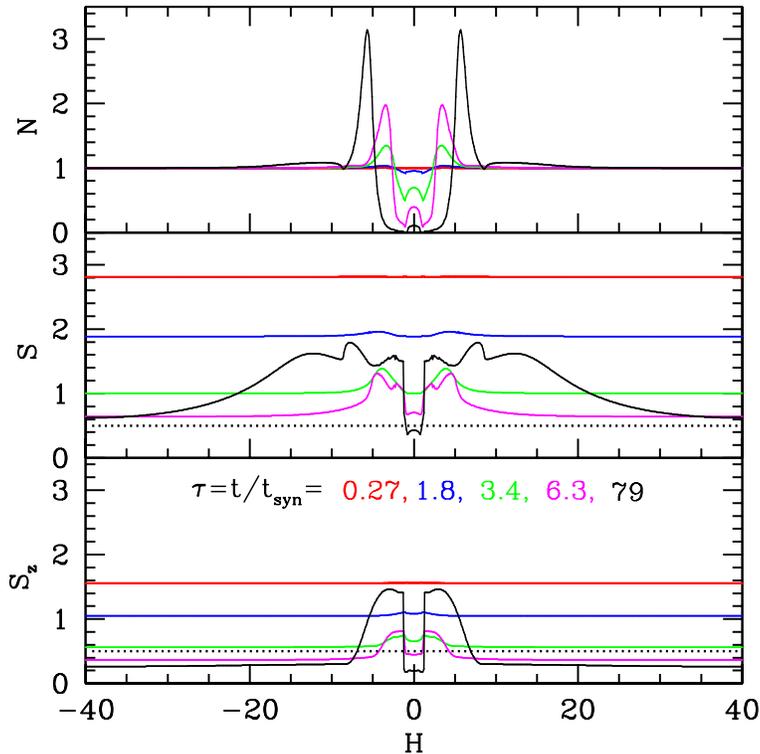}
\caption{Time evolution of the spatial distributions of planetesimal 
surface density $N$ (normalized by its value at infinity), 
eccentricity and inclination dispersions $S$ and $S_z$
scaled by the embryo's Hill radius for $N a r_H=400, p=0.004, {\cal M}_0=
10, S_{0}=0.5, S_{z 0}=0.5$. Curves of different color represent 
the snapshots of the distributions at different moments of time marked with 
corresponding color (also marked on the evolutionary track in Figures
\ref{fig:evol_2.6_2.4}, \ref{fig:mass_2.6_2.4}, \& 
\ref{fig:vel_2.6_2.4} with starred dots). The dotted line shows the initial
state of the disk.  See text for more details.}
\label{fig:time_ev}
\end{figure}

\begin{figure}[t]
\vspace{10.5cm}
\includegraphics{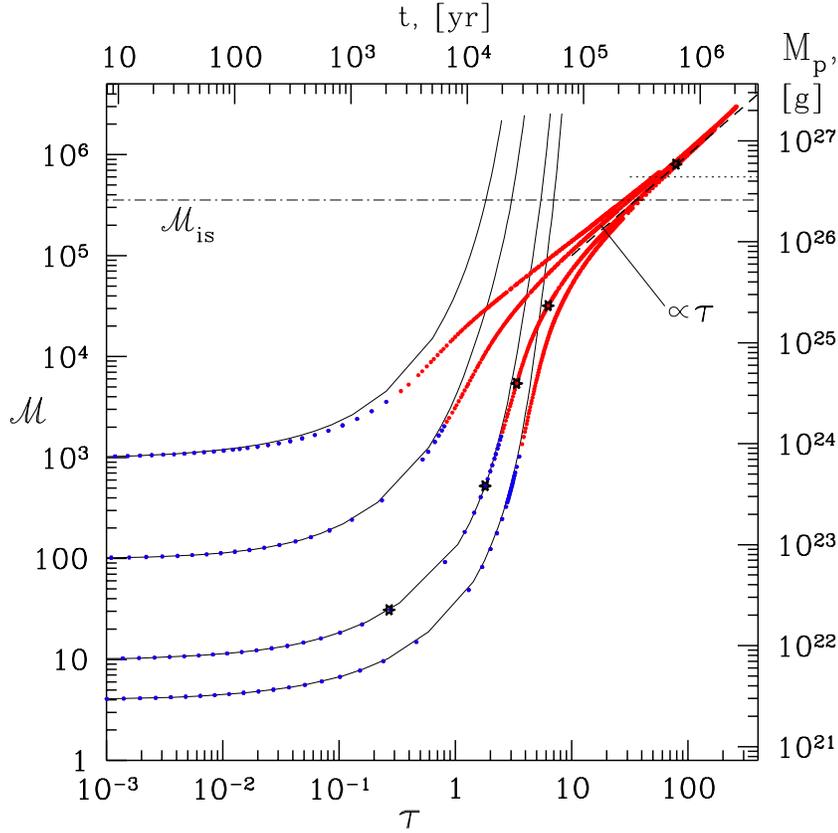}
\caption{Growth of the embryo's mass in the case $N a r_H=400$, $p=0.004$
for different initial conditions. Mass 
${\cal M}$ as a function of time $\tau$
is plotted by dots. The thin solid lines display corresponding
runaway curves obtained by neglecting embryo's dynamical effects.
The dimensional values of time $t$ and mass 
of the embryo $M_e$ are obtained in 
the same way as in Figures \ref{fig:evol_2.6_2.4}. The dot-dashed line 
represents the isolation mass in the shear-dominated regime 
defined by (\ref{eq:isol_mass}) and by
(\ref{eq:isol_mass_dd}) in the dispersion-dominated regime (see text).
The dashed line has a slope of unity and shows the asymptotic behavior of 
${\cal M}$ [see equation (\ref{eq:m_tau_eredict})].}
\label{fig:mass_2.6_2.4}
\end{figure}

\begin{figure}[t]
\vspace{10.5cm}
\includegraphics{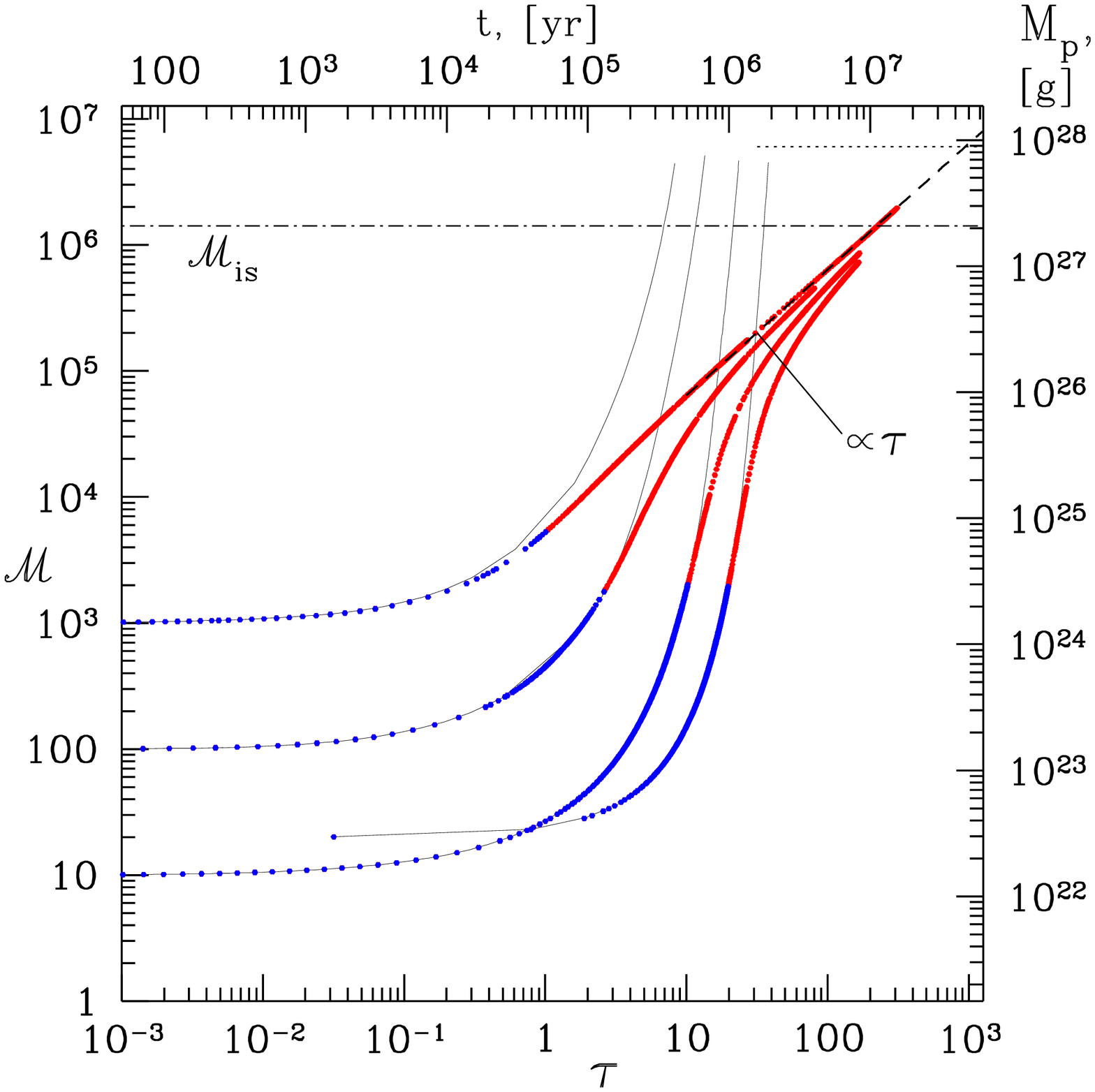}
\caption{The same as Figure \ref{fig:mass_2.6_2.4} but for 
$N a r_H=10^3$, $p=10^{-3}$.}
\label{fig:mass_3_3}
\end{figure}

\begin{figure}[t]
\vspace{10.5cm}
\includegraphics{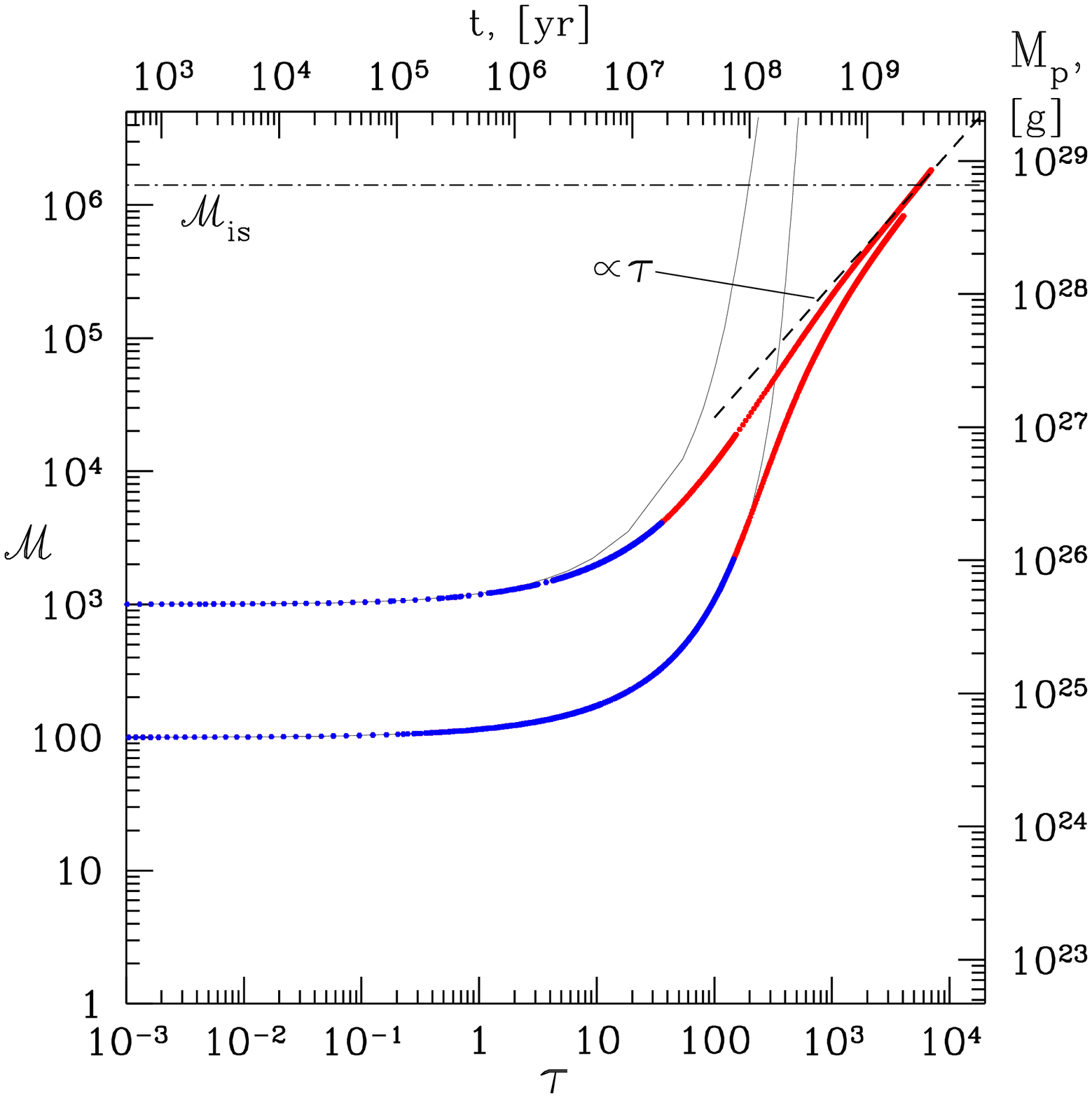}
\caption{The same as Figure \ref{fig:mass_2.6_2.4} but for 
$N a r_H=10^3$, $p=10^{-4}$.}
\label{fig:mass_3_4}
\end{figure}

\begin{figure}[t]
\vspace{10.5cm}
\includegraphics{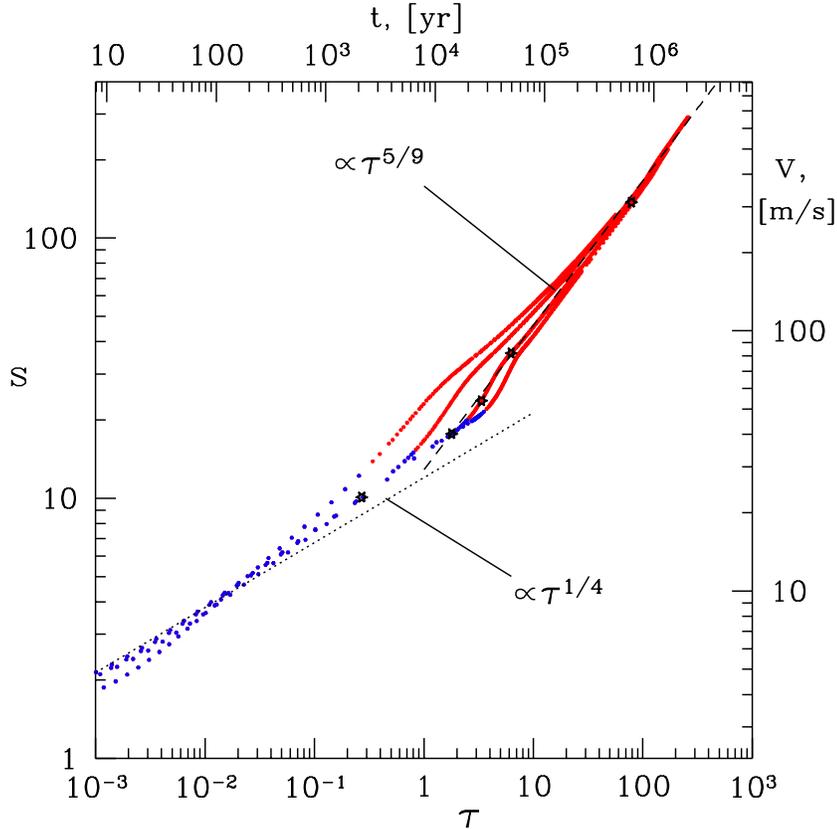}
\caption{Growth of the disk velocity dispersion $s$ in the heated zone as
a function of time $\tau$ for $N a r_H=400$, $p=0.004$. Sequences of dots
display $s(\tau)$ for different initial conditions. The dotted line has
a slope of $1/4$ and is intended to illustrate the initial growth
of $s$ due to the planetesimal-planetesimal scattering alone 
[see (\ref{eq:sols_el})]. The dashed line has a slope of $5/9$ 
and shows the asymptotic behavior of $s$. }
\label{fig:vel_2.6_2.4}
\end{figure}

\begin{figure}[t]
\vspace{10.5cm}
\includegraphics{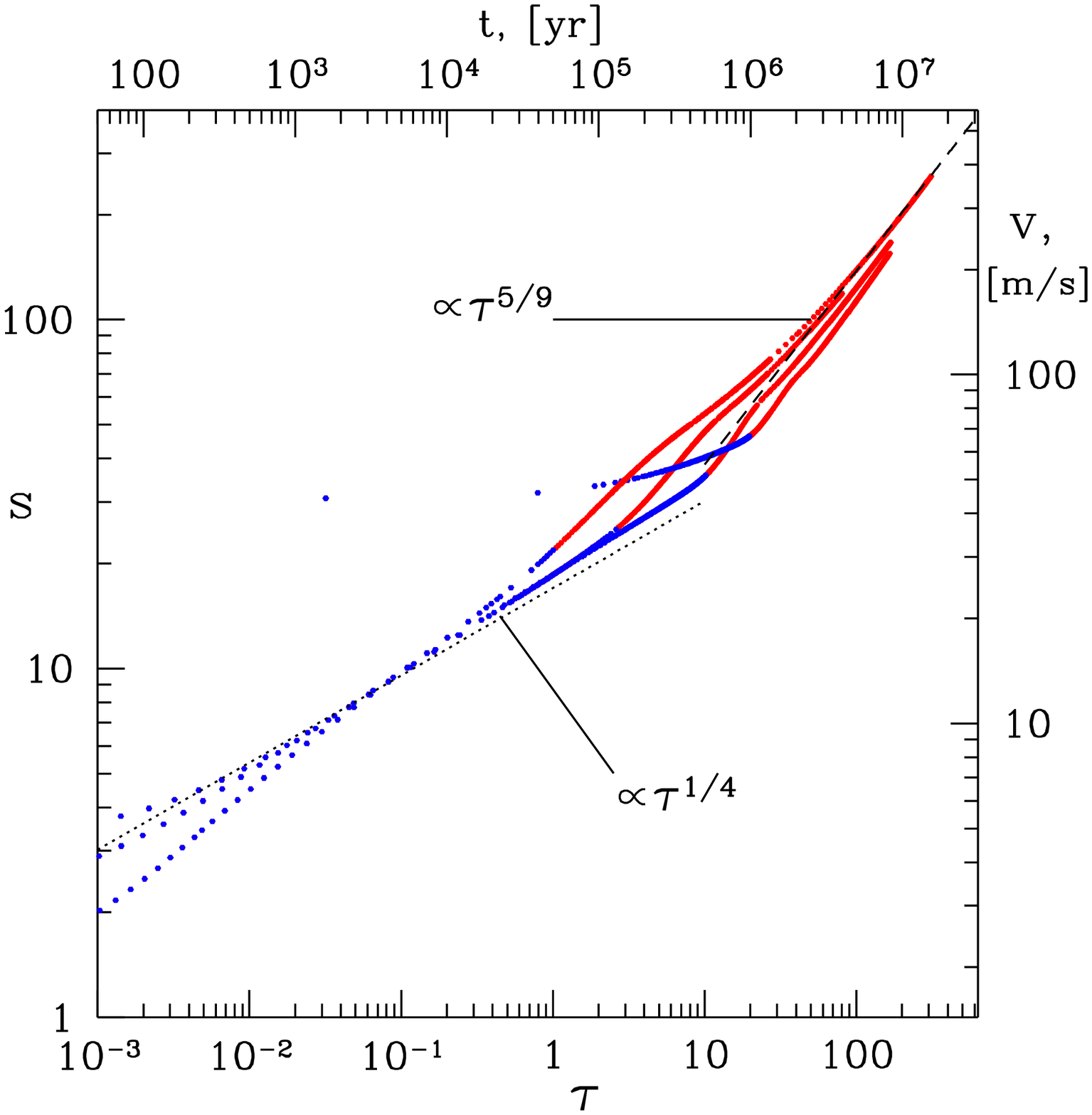}
\caption{The same as Figure \ref{fig:mass_2.6_2.4} but for 
$N a r_H=10^3$, $p=10^{-3}$.}
\label{fig:vel_3_3}
\end{figure}

\begin{figure}[t]
\vspace{10.5cm}
\includegraphics{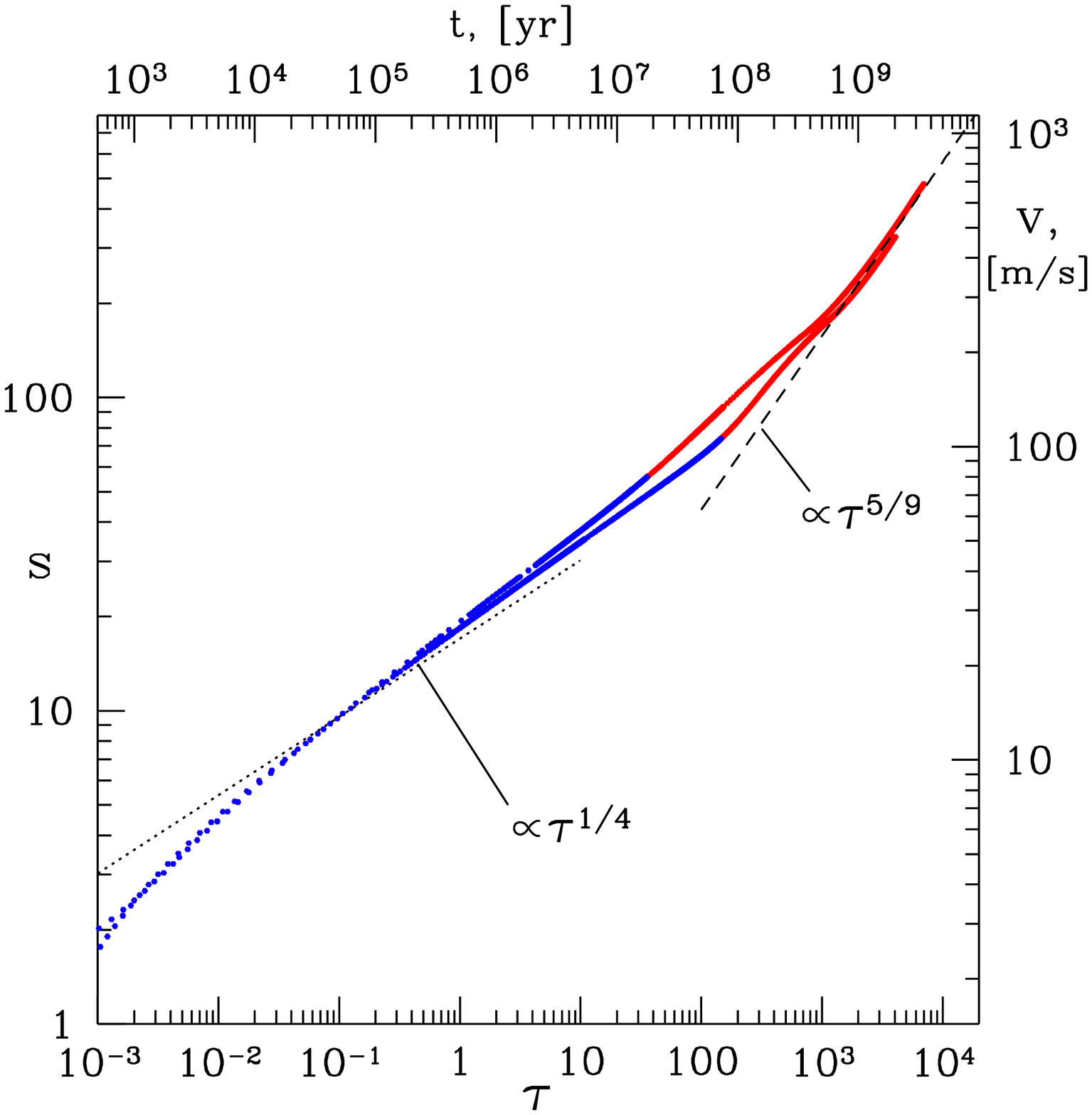}
\caption{The same as Figure \ref{fig:mass_2.6_2.4} but for 
$N a r_H=10^3$, $p=10^{-4}$.}
\label{fig:vel_3_4}
\end{figure}

\begin{figure}[t]
\vspace{10.5cm}
\includegraphics{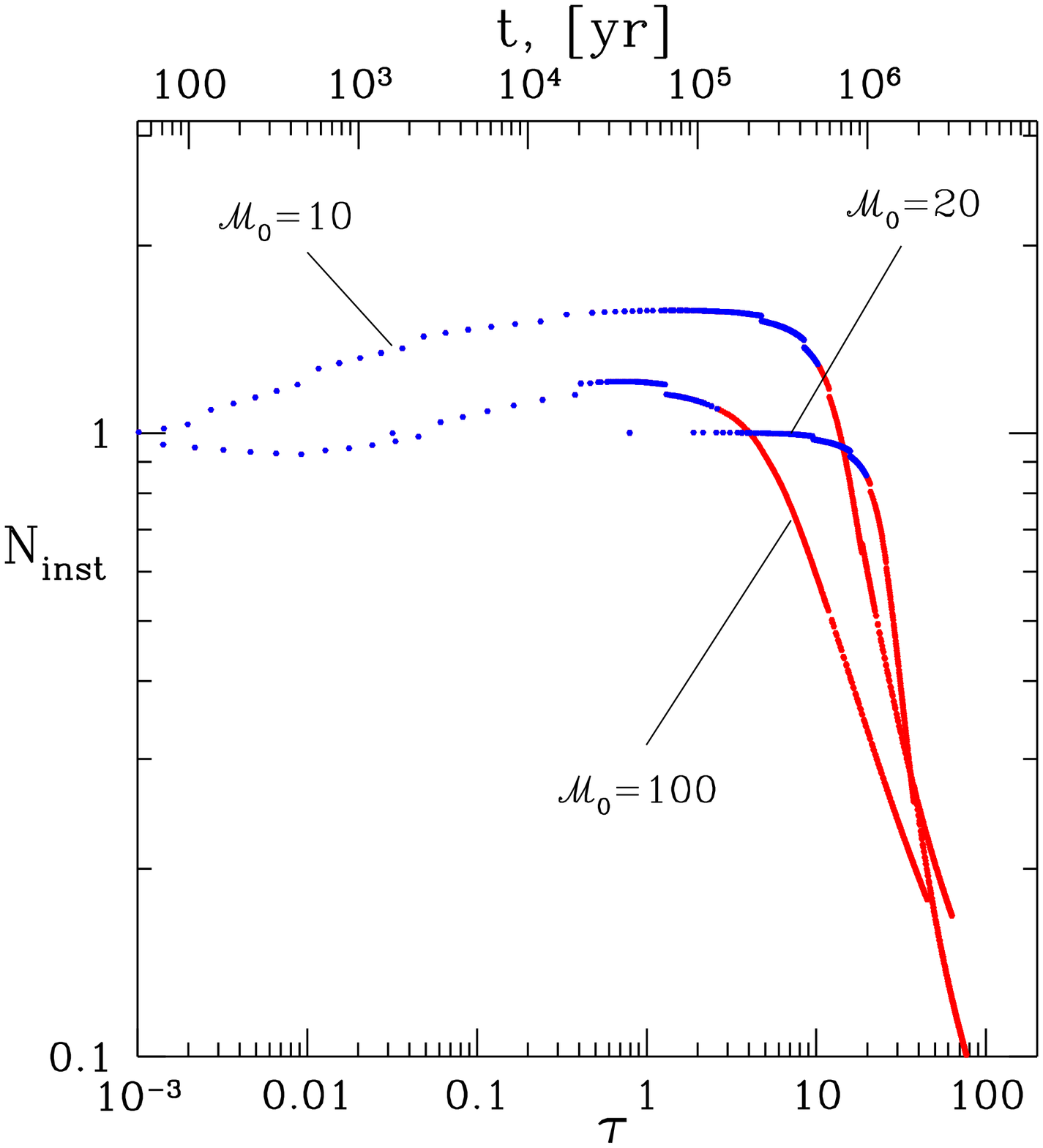}
\caption{Instantaneous surface density of planetesimals which can be 
accreted by the embryo $N_{inst}$ as a function of time. 
Curves for several 
different values of initial embryo's mass are shown (labeled on the figure).
Note the sharp drop of $N_{inst}$ after embryo starts to dominate the 
 dynamics of the heated zone.}
\label{fig:ninst_3_3}
\end{figure}


\subsection{General description of evolution.}
\label{subsect:gen_evol}


The results of the numerical solution of the evolution equations are 
presented in Figures \ref{fig:evol_2.6_2.4}-\ref{fig:vel_3_4}.
We use the following general notation when plotting evolutionary tracks
of different quantities (${\cal M}$ or $s$): 
tracks corresponding to the regime where planetesimal-planetesimal
scattering dominates the disk dynamics are plotted
in blue,
while those in the regime dominated by embryo-planetesimal scattering
are shown with
red.
The separation between these two regimes is determined based on 
Figures \ref{fig:vel_2.6_2.4}-\ref{fig:vel_3_4} (see below).
We also plot on each Figure (on the right and top borders) 
the numerical values
of physical variables $M_e$ (mass of the embryo) and $v$ (linear velocity
of planetesimal epicyclic motion) corresponding to their dimensionless analogs
${\cal M}$ and $s$. In doing this we take into account the 
aforementioned property [see discussion after (\ref{eq:nar_H}) 
and ({\ref{eq:p}) of $N a r_H$ and $p$ to unambiguously specify
$a$ and  $m$ when $\rho$ and $M_c$ are fixed.
Also, where appropriate we plot not only the evolutionary tracks 
including the effects of embryo's gravity on the disk dynamics 
(``active'' embryo, colored dots) 
but also the 
curves calculated assuming that embryo does not affect the disk dynamics
(thin solid lines); in other words we 
allow the disk to stay completely 
homogeneous and to heat up only due to the 
planetesimal-planetesimal scattering,
while still allowing the embryo to grow (``passive'' embryo). 
This would correspond to
the outcome of conventional coagulation simulations
(in our problem setting) which do not take
embryo-induced disk inhomogeneities into account.

In Figures \ref{fig:evol_2.6_2.4}-\ref{fig:evol_3_4} we plot the 
evolution of the embryo-disk system in coordinates $s$-${\cal M}$
for different values of $N a r_H$ and $p$. This can be compared
to the schematic calculations in  Figure \ref{fig:regions_mass}. 
Each of the plots displays several  
evolutionary tracks shown by sequences of dots. These tracks  usually 
start with moderate values of ${\cal M}_0$ (from $10$ to $10^3$) and rather
small values of $s_0$, so that the embryo-disk interaction can start either 
in the shear- or in the dispersion-dominated regimes.
A brief inspection of  Figures 
\ref{fig:evol_2.6_2.4}-\ref{fig:evol_3_4} shows that the coupled evolution of
the embryo-disk system can be naturally split into three distinct 
regimes. In the first regime the mass of the embryo stays almost 
constant\footnote{This happens because the timescale of the runaway growth 
is longer than the timescale of the disk heating.} and
only the velocity dispersion in the disk grows. After a short intermediate
stage when both $s$ and ${\cal M}$ evolve at comparable rates, the embryo 
 grows faster than disk heats up because of the 
fast runaway accretion. 
At some intermediate point during this stage the embryo begins to  
dominate the dynamical evolution of the planetesimal disk
(blue dots change to red dots). As a result the evolutionary track 
gradually departs from the runaway path and accretion proceeds 
at a slower rate. Finally, the system settles onto the asymptotic regime 
in which embryo's mass grows as a power law of the velocity dispersion 
in the heated zone $s$. 

Figure \ref{fig:time_ev} provides more insight into what is going on in 
the disk by plotting the time evolution of the spatial distributions
of planetesimal surface density $N$ and eccentricity and inclination 
dispersions $S$ and $S_z$ [scaled by $R_H$, see (\ref{eq:new_vels})].
These distributions are plotted for an evolutionary track with
$N a r_H=400, p=0.004$ and initial values of mass and velocities
${\cal M}_0=10, S_{0}=S_{z 0}=0.5$. We also plot the locations of the
points at which the snapshots of spatial distributions are taken 
in Figures \ref{fig:evol_2.6_2.4}, \ref{fig:mass_2.6_2.4}, and 
\ref{fig:vel_2.6_2.4} (black stars on 
the corresponding evolutionary track). One can see that initially the disk
 heats up very quickly: $S$ grows from $0.5$ to $2.8$, $S_z$ from $0.5$
to $1.6$ on a rather short timescale. 
The ratio of $S_z/S$ is close to $0.55$, which is  characteristic for 
homogeneous Keplerian disks (Ida \etal 1993; Stewart \& Ida 2000). 
One can see that the embryo's effects are 
negligible and disk is very uniform (red curve). After that rapid 
runaway accretion of the embryo leads to the decrease of $S$ and $S_z$ because
growth of $R_H$ surpasses the increase of $s$ and $s_z$. 
Disk inhomogeneities are rather small while disk dynamics is 
still dominated by the mutual planetesimal encounters (blue curve). 
However, as soon as the 
embryo takes over the control of the disk heating (green curve) deviations
from the uniform state become important:
sharp features appear in the distributions of $S$ and $S_z$ and the
embryo carves out a gap in surface density of planetesimal guiding centers 
around itself (results analogous to those of Papers I \& III). 
After the intermediate stage
(magenta curve), the disk finally settles into the asymptotic regime
(black curve) and the difference between the planetesimal velocities in the
heated zone and far from the embryo steadily grows with time.
The detailed shapes of the spatial distributions of different quantities 
are almost invariant in the asymptotic regime --- 
we would obtain the same final profiles if we were to start with
a different set of ${\cal M}_0, S_{e 0}, S_{i 0}$. Note that
far away from the embryo, the disk always stays uniform and $S_z/S$ 
is always near $0.55$\footnote{This ratio of $S_z/S$ eventually falls
in the very late 
stages of embryo-disk evolution, because eccentricity
is rather effectively excited by the distant encounters
(see e.g. Hasegawa \& Nakazawa 1990).}.

It is interesting that for a rather long period of time disk 
evolution in the heated zone takes place 
almost in the shear-dominated regime.
This happens because the runaway growth of 
the embryo's mass is much faster than
the growth of random velocities in the disk.
This switches embryo-planetesimal
interactions from the dispersion-dominated into the intermediate
velocity regime.
Some minor details of the spatial distributions in this regime might be 
not real because our analytical 
apparatus developed in Paper III is not very
accurate in the intermediate velocity regime; however the overall description
of the disk evolution should be robust.

We have mentioned in \S \ref{subsect:regime_separation_all} that
planetesimal-planetesimal scattering leads not only to the excitation
of planetesimal random motions but also to the spatial redistribution of  
their kinetic energy and surface density (see also Paper II). This is 
equivalent to the action of the effective viscosity in the planetesimal 
disk which tries to smooth out the gradients of the surface density $N$
and velocity dispersions $S$ and $S_z$. 
As we have mentioned in \S \ref{subsect:regime_separation_all} the magnitude
of the effect produced by the viscosity is typically the same as that of
the direct excitation of planetesimal velocities by 
planetesimal-planetesimal scattering. Planetesimal viscosity 
is very large initially when planetesimal velocities are small. As a result
the spatial profiles of $N$, $S$ and $S_z$ are very smooth 
(red or blue curves). Later on gravitational effects of the 
embryo increase as well as the planetesimal velocities; this acts to
diminish 
the role of viscosity --- spatial distributions start to exhibit 
sharp features driven by the embryo's gravity (green and magenta curves). 
Finally, deep in the asymptotic regime (black curve)
where $s$ is very 
large viscosity is very small: it cannot even smooth out strong 
gradients of $S$ and $S_z$ at the edges of the region occupied 
by the planetesimals on the horseshoe orbits (planetesimal heating inside 
this zone is very weak). This justifies the validity of the concept 
of the ``heated'' zone: when embryo dominates nearby disk dynamics, the
planetesimal viscosity which couples the excited region 
to the distant parts of the 
disk becomes relatively unimportant, and the heated region can be considered
as dynamically isolated from the rest of the disk. 

The general picture of the embryo-disk evolution outlined in 
Figures \ref{fig:evol_2.6_2.4}-\ref{fig:time_ev} 
is in excellent agreement with the predictions of 
our simple qualitative analysis presented in \S \ref{sect:scaling}. 
We have  attempted to provide some quantitative comparisons
by fitting analytical results of \ref{subsect:regime_separation} 
and \ref{subsect:regime_separation_all} 
to the boundaries of various regions in Figures 
\ref{fig:evol_2.6_2.4}-\ref{fig:evol_3_4}. For example, using the transition
between the regions where embryo and planetesimals separately dominate
the disk dynamics (where the color of the dots changes in our plots) 
we can estimate the approximate position
of a critical point $(s_\star,{\cal M}_\star)$ defined by 
(\ref{eq:m0_s0}):
\begin{eqnarray}
{\cal M}_\star\approx 40~(N a r_H)^{3/5},~~~s_\star\approx
3.4~(N a r_H)^{1/5}.
\label{eq:star_eredict}
\end{eqnarray}
In making this estimate we have neglected the variation of logarithmic 
factor with $N a r_H$ and treated it as a constant which is absorbed in 
the coefficient in (\ref{eq:star_eredict}). We forced this point to lie
on the line ${\cal M}=s^3$ (thin solid line) separating shear- and 
dispersion-dominated regimes of the embryo-planetesimal interaction.
In the same spirit we have found that the position of the separatrix
between the regions of planetesimal and embryo dominance over the disk 
dynamics ${\cal M}_{dd}(s)$ [see equation (\ref{eq:bound_dd})] can be roughly
fitted by
\begin{eqnarray}
{\cal M}_{dd}(s)\approx \left[\frac{{\cal M}_\star^{5/3}}
{\ln\Lambda_e}\right]^{1/2}s^{1/2},~~~\Lambda_e=1+\frac{s^3}{{\cal M}_\star}.
\label{eq:m_dd_eredict}
\end{eqnarray}
This relation is shown by the long-dashed curve; one can see that within
a factor of several it correctly predicts the transition from 
planetesimal to embryo's dominance of the disk dynamics 
although the agreement gets worse
as one moves towards the shear-dominated regime [the continuation of
this separatrix into the shear-dominated region is drawn using 
(\ref{eq:bound_sd})]. We have kept the logarithmic dependence on $s$ in
(\ref{eq:m_dd_eredict}) since it slightly improves agreement with the 
numerical results. Our estimate of ${\cal M}_{dd}$ agrees
with the corresponding condition suggested by Ohtsuki \& Tanaka (2002),
see their equation (39). They have the same functional 
dependencies\footnote{Note that the estimate of Ohtsuki \& Tanaka 
(2002) mistakenly lacks a factor $\ln\Lambda_e$ 
(meaning that embryo-planetesimal 
interaction is in the dispersion-dominated regime) while still 
retaining $\ln\Lambda_e$.} and numerical coefficients of the same order
of magnitude.

One can see that the evolutionary tracks which were calculated without 
taking account of the 
embryo's effect on the disk dynamics (thin solid curves)
coincide very well with those which include embryo's gravity while the disk 
evolution is dominated by planetesimal-planetesimal scattering. This is
expected because when ${\cal M}<{\cal M}_{dd}$ the embryo cannot perturb
the disk surface density or appreciably affect planetesimal velocities 
even in its own vicinity. As a result the disk stays homogeneous and heats up
only by planetesimal encounters --- exactly the situation for which the 
curves with a ``passive'' embryo were computed. However, as soon as the 
embryo takes over the control of the disk dynamics the evolutionary track
with an ``active'' embryo departs from that of a
``passive'' embryo in agreement
with equation (\ref{eq:relat_em}): disk heating speeds up under the action
of the embryo's gravity while the growth of embryo's mass slows down 
as a result of the weaker gravitational focussing of incoming planetesimals
and, correspondingly, smaller collision cross-section.

Finally, in the late stages of its evolution the embryo-planetesimal system
settles onto an attractor. There the dependence of
${\cal M}$ on $s$ can be fitted by a power law 
with the index $9/5$ [see equation (\ref{eq:m-em-separation})]. 
For a given $N a r_H$ and $p$ all evolutionary tracks independent of
their starting ${\cal M}_0$ and $s_0$ end up on the same curve
exactly as envisaged by our analysis in \S 
\ref{subsect:regime_separation_all}. Using numerical results for different 
$N a r_H$ and $p$ this curve can be approximated by
[see (\ref{eq:m-em-separation})]
\begin{eqnarray}
{\cal M}_{M-em}(s)\approx 90~[p(N a r_H)]^{3/5} s^{9/5}.
\label{eq:m_em_eredict}
\end{eqnarray}
This relation is shown by the short-dashed line in 
Figures \ref{fig:evol_2.6_2.4}-\ref{fig:evol_3_4}. 
It is clearly an attractor to which all possible initial configurations
finally converge. At some point this solution reaches the isolation mass
${\cal M}_{is}$ which is displayed by the short dash-dotted line ---
given by (\ref{eq:isol_mass}) in the shear-dominated regime and
by (\ref{eq:isol_mass_dd}) in the dispersion-dominated regime. It is 
important to stress 
that ${\cal M}_{is}$ is reached in the asymptotic regime of the
embryo's evolution and not on the runaway stage. This has profound 
implications for the formation timescale of planetary embryos
(see \S \ref{sect:discussion}).

We have also checked the agreement between the predictions of 
\S \ref{subsect:regime_separation_all} and our numerical results for the 
location of the separatrix between the regions 
where the embryo grows faster or slower than planetesimals heat up the disk.
This is the line where ${\cal M}={\cal M}_{M-pl}(s)$ or $T_{pl}=T_{em}$ 
(see \S \ref{subsect:regime_separation_all}). 
In numerical calculations we set this transition at the point
where the slope of evolutionary tracks in $s$-${\cal M}$ coordinates
is equal to 1, i.e. where $d\ln{\cal M}/d\ln s=1$. The resulting
curves for different $N a r_H$ and $p$ are well fitted by 
[see equation (\ref{eq:m-pl-separation})]
\begin{eqnarray}
{\cal M}_{M-pl}(s)\approx\frac{0.15}{(ps^2)^3},
\label{eq:m_el_eredict}
\end{eqnarray}
where the logarithmic factor $\ln\Lambda_e$ is 
again absorbed into a constant coefficient.

Note that the values of $p$ and $N a r_H$ we have used in our 
calculations typically do not lead to the crossing
of ${\cal M}_{M-pl}(s)$ with ${\cal M}_{M-in}(s)$ in the 
dispersion-dominated regime [meaning that $p>p_0$, see equation 
(\ref{eq:p_0_def})]. Nevertheless, our numerical results are still in good
concord with the 
analytical results of \S \ref{subsect:regime_separation_all} 
which were derived assuming that these curves do cross in the 
dispersion-dominated regime [i.e. $p<p_0$]. This is explained by the 
argument put forward in the end of \S \ref{subsect:regime_separation_all}: 
values of $N a r_H$ and $p$ that we have considered are not very extreme 
($N a r_H$ is not large enough and $p$ is not small enough) and,
as a result, evolutionary tracks do not explore regions of 
$s$-${\cal M}$ space characteristic for the shear-dominated case. They 
only dwell for some time in the intermediate zones where transitions from one 
asymptotic regime to the other occur. Of course, if one were to explore e.g. 
larger values of $N a r_H$ some differences from the case studied in 
\S \ref{subsect:regime_separation_all} would appear. We expect though 
that this would not change our principal conclusions.


\subsection{Time dependence of ${\cal M}$ and $s$.}
\label{subsect:M_and_s}


In Figures \ref{fig:mass_2.6_2.4}-\ref{fig:mass_3_4} 
we plot the dependence of embryo's mass on time for the
evolutionary tracks shown in Figures 
\ref{fig:evol_2.6_2.4}-\ref{fig:evol_3_4}. Here one can see better
that disk evolution with the ``passive'' embryo (thin solid curves) really
leads to a runaway --- mass of the embryo grows to extremely large values
in a finite time. We have found that mass growth on this runaway stage can be 
approximately described by equation (\ref{eq:sols_el}) with $\tau_{run}$
roughly (within a factor of 2) given by
\begin{eqnarray}
\tau_{run}\approx\frac{0.08}{{\cal M}_0^{2/3}(N a r_H) p^2}.
\label{eq:tau_run_eredict}
\end{eqnarray}
Evolutionary paths corresponding to the more realistic case of
an ``active'' embryo depart from the runaway tracks when the embryo starts 
dominating disk dynamics in its vicinity, just as in Figures 
\ref{fig:evol_2.6_2.4}-\ref{fig:evol_3_4}. They smoothly switch to
a behavior where ${\cal M}$ is a power law of time [similar behavior was obtained
in Weidenschilling \etal (1997)]. Although this
is consistent with the results of our qualitative analysis the exact
value the power law index does not coincide with the
value of $9/7$ 
predicted by equation (\ref{eq:sols_em}). In fact it is closer
to 1 so that the embryo's mass grows linearly with time. We have 
found that the time dependence of mass can be roughly described by
the following relation:
\begin{eqnarray}
{\cal M}(\tau)\approx 1300~ (N a r_H)^{1.63}p^{1.4}\tau.
\label{eq:m_tau_eredict}
\end{eqnarray}

There is a simple 
explanation of this apparent discrepancy: when the embryo dominates the 
disk heating it induces strong nonuniformities in the distribution of
the planetesimal surface density around it. As a result the instantaneous
surface density of planetesimals at the embryo's location which 
determines its accretion rate decreases with time. We illustrate this 
point in Figure \ref{fig:ninst_3_3} where we plot the instantaneous surface 
density of planetesimals near the embryo $N_{inst}$
normalized by its initial value 
as a function of time $\tau$ for $N a r_H=10^3$, 
$p=10^{-3}$ and several different values of initial mass ${\cal M}_0$.
Values of $N_{inst}$ are computed from spatial profiles of $N$
and $S$ using the corresponding conversion found
in Paper II; in calculating $N_{inst}$ we exclude the contribution of
the horseshoe orbits because they cannot approach the embryo very closely 
and be accreted. One can see that $N_{inst}$ begins to vary strongly 
only after the embryo starts dominating the 
disk dynamics (where the color of dots 
changes to red); in this regime $N_{inst}$ can drop by a factor of 10 
or more because of the spatial redistribution of planetesimals in the disk. 
This is equivalent to a decrease of $N a r_H$ which 
would otherwise be constant. As a result, the accretion rate 
diminishes and embryo's mass grows slower than equation 
(\ref{eq:sols_em}) would predict.
The small variations of $N_{inst}$ that occur in the stage 
when planetesimal dynamics is
determined only by planetesimal scattering (blue dots) are explained by
our exclusion of planetesimals on horseshoe orbits\footnote{In this 
regime the planetesimal disk is almost homogeneous; if we did not exclude
planetesimals on the horseshoe orbits $N_{inst}$ would be exactly constant.}:
the spatial extent of the horseshoe region depends on the eccentricity and 
inclination of the planetesimals (see Paper III for details)
which vary with time; this causes $N_{inst}$ to change.

Our simple theory presented in  \S \ref{subsect:regime_separation_all}
does not account for the local decrease of the planetesimal  surface density
as ${\cal M}$ the planetesimals are accreted onto the embryo. 
This is a considerable drawback which requires a more
sophisticated treatment. We will not pursue this 
subject in this paper, merely note on its importance. Luckily it does not
affect the general picture of the evolution of embryo-disk system described
in \S \ref{subsect:regime_separation_all}, up to the isolation mass
${\cal M}_{is}$. 

Note a very strong increase of the embryo's growth timescale 
in Figures \ref{fig:mass_2.6_2.4}-\ref{fig:mass_3_4} as one
moves towards the outer edge of the nebula. The dimensionless timescale
(in units of $\tau$) grows because it strongly depends on parameter $p$
characterizing embryo's accretion cross-section and $p$ diminishes when
the distance from the central star $a$ increases. The timescale
measured in physical units is additionally lengthened by the growth
of the planetesimal synodic period $t_{syn}$
with $a$: $t_{syn}\propto a^{3/2}$. As a result the most
favorable conditions for the embryo's growth exist in the inner parts 
of the nebula. We will return to this issue in \S \ref{sect:discussion}. 

In Figures \ref{fig:vel_2.6_2.4}-\ref{fig:vel_3_4} 
we plot the dependence of $s$ on time in the same way as we did it for
${\cal M}$. Initially (when planetesimal encounters control disk
dynamics) one would expect $s$ to behave in accordance with 
equation (\ref{eq:sols_el}) which is exhibited by dotted line. Deviations
from this power-law solution are caused by additional logarithmic 
dependence present in equation (\ref{eq:sols_el}). At some point evolution
switches to the embryo-dominated regime. We have used
Figures \ref{fig:vel_2.6_2.4}-\ref{fig:vel_3_4} to determine the 
locations of the corresponding
transition points for each track by approximately picking
the point at which there is a break in the $s(\tau)$ dependence. As we have 
discussed above, this choice of the transition points agrees well
with other characteristic indicators e.g. splitting
of ${\cal M}(\tau)$ for ``passive'' and ``active'' embryo tracks in 
\ref{fig:mass_2.6_2.4}-\ref{fig:mass_3_4}. When $\tau$ becomes very large 
system evolves  in the asymptotic regime and 
$s$ grows as a power-law of $\tau$ with index close to $5/9$. This 
particular dependence and the fact that ${\cal M}(\tau)\propto \tau$ in 
the asymptotic regime conspire to produce the dependence of
${\cal M}(s)$ which is very close to that predicted analytically 
[see (\ref{eq:m-em-separation}) and (\ref{eq:m_em_eredict})].  
The fact that the power-law index of $s(\tau)$ is different from 
$5/7$ predicted by (\ref{eq:sols_em}) is again explained by the drop of 
$N_{inst}$ in the asymptotic regime: since embryo
grows slower that expected it heats up planetesimal population nearby 
less efficiently than our analysis of 
\S \ref{subsect:regime_separation_all} predicts. Nevertheless, by the
end of our calculations planetesimal velocity dispersion in the heated zone 
exceeds the random velocity in the distant part of 
the nebula typically by a factor of several. As expected, the gravitational
influence of the embryo considerably speeds up the dynamical evolution
of the planetesimals in the heated zone. This behavior is in general agreement with 
the results of multi-zone simulations by Weidenschilling \etal (1997).


\section{Discussion.}
\label{sect:discussion}



\subsection{Results and their applications.}
\label{subsect:applications}


The most important result of our investigation is the elucidation
of the role played by the coupling between the {\it nonuniform} 
disk dynamics and the embryo's mass growth. We have found in 
agreement with the previous 
investigations (Ida \& Makino 1993; Tanaka \& Ida 1997; Paper I) that 
the dynamical excitation of the planetesimal disk by the gravity of 
growing protoplanetary embryo dramatically changes the simple runaway 
picture of planetary growth, which is derived for 
homogeneous planetesimal disks.
Using our results we can also study quantitatively the question of 
how long is required for the embryo to form?

As we have pointed out in \S \ref{sect:numerical},
embryo accretes all the material within its feeding zone and 
reaches the isolation only after it settles into the asymptotic 
evolutionary regime (and not on the runaway stage). 
The time that the embryo spends
on this asymptotic stage dominates the final timescale of 
its evolution. Indeed, for all the values of $N a r_H$ and $p$ explored
in \S \ref{sect:numerical} the runaway timescale is usually shorter by a factor
of several than the time needed for the embryo to reach isolation
in the asymptotic stage.
This might indicate that the estimates of the embryo formation timescale 
produced by the conventional coagulation simulations neglecting the 
effects of inhomogeneous disk evolution could be off by a 
substantial factor.

For example, in the case $N a r_H=400$, $p=0.004$ typical for the 
planet formation in the terrestrial planet region of MMSN, the 
time needed for
``passive'' runaway embryo (thin solid curve) 
with $M_e/m=10$ to reach the isolation mass 
$\approx 3\times 10^{26}$ g is
$\approx 4\times 10^4$ yr; growth to the same mass for an 
``active'' embryo that affects the disk would take $\approx 3\times 10^5$ yr 
(see Figure \ref{fig:mass_2.6_2.4}) ---
almost a factor of $10$ increase! This difference is even more striking
for larger initial $M_e/m$ for which runaway timescales are very short.

The very long time needed for the embryo's growth which 
arises in our calculations exacerbates the timescale 
 problem for the growth of the rocky cores of
the giant planets,  because this process usually takes at least as long
as the gaseous nebula is supposed to last: at $3.6$ AU in the MMSN
(a situation described by $p=10^{-3}$ and $N a r_H=10^3$ in our 
computations) the growth of $M_e$ to $2\times 10^{27}$ g [the isolation mass
given by (\ref{eq:isol_mass})] takes $10^7$ yr, while reaching
$9\times 10^{27}$ g [isolation mass 
given by (\ref{eq:isol_mass_dd}) which accounts for the expansion of the 
feeding zone due to the excitation of planetesimal random motions]
would take $5\times 10^7$ yr. Not only is this timescale is too long, it
is also not clear that the isolation mass is large enough to trigger
the core instability which would allow giant planets to accrete their 
huge gaseous mass (Mizuno 1980; Bodenheimer \& Pollack 1986). 
Post-isolation chaotic dynamical evolution of the embryos (Chambers 2001) 
is the most likely solution of the
latter problem, but it would require even more time for building up 
massive gas-accreting planetary cores. Note that the existence of 
this post-runaway chaotic evolutionary stage would not be a 
problem for the formation of terrestrial planets 
because (1) timescales are shorter in the inner parts of the nebula
(see Figure \ref{fig:mass_2.6_2.4}),
and (2) terrestrial planets did not accrete large amounts of gas which
might 
indicate that they have formed after the gaseous nebula dispersed
(which eliminates the time constraint related to the nebular dispersal). 

It is true 
that in our calculations the isolation mass is usually larger
than ${\cal M}_{is}$ reached by the embryos growing in the 
runaway regime in homogeneous disks:
in the runaway stage ${\cal M}_{is}$ is given by (\ref{eq:isol_mass}) since 
it is typically reached in the shear-dominated regime (see Figures
\ref{fig:evol_2.6_2.4}-\ref{fig:evol_3_4}) while for the ``active''
embryos ${\cal M}_{is}$ is increased as a result of the strong 
excitation of planetesimal random velocities by the embryo, see equation 
(\ref{eq:isol_mass_dd}).
This simplifies the formation 
of massive rocky cores (at $3.6$ AU ${\cal M}_{is}$ is increased by almost 
a factor of $5$, see above). 
Unfortunately, the time required to reach larger ${\cal M}_{is}$ 
is longer as well.

The timescale issue becomes even more challenging for the growth 
of the outer giant planets if they  
formed at their present locations\footnote{It was suggested by 
Thommes \etal (2002) that
Uranus and Neptune were formed between the orbits of Jupiter and Saturn and 
were then scattered into their present orbits by these giant planets.}:
typical formation timescales are $\sim 10^9$ yr 
(see Figure \ref{fig:mass_3_4}) which is much longer than the timescale
of the nebular dissipation (Hollenbach \etal 2000). 
One possibility to alleviate this problem would be to increase the 
surface density of planetesimals; this is a reasonable assumption 
beyond the so-called ``snow line'' (Hayashi 1981; Sasselov \& Lecar 2002) 
where the condensation of ice
takes place (at $1-3$ AU).

The fact that the embryo's growth does not stop when the 
embryo-planetesimal
interaction starts to dominate disk dynamics and opens a gap 
is quite interesting. Previous estimates of the gap opening
mass (Paper I; Ohtsuki \& Tanaka 2002) only
considered an embryo with a fixed mass and allowed plenty of time 
for  the disk evolution. Under these conditions a gap would 
form in the planetesimal disk 
around the embryo when $T_{em}\approx T_{pl}$ and
${\cal M}\approx {\cal M}_{dd}(s)$; the 
instantaneous accretion rate of the embryo
(if it were able to increase its mass) would be strongly affected
even at this rather small value of ${\cal M}$.
This would strongly slow down embryo's accretion at
${\cal M}\approx {\cal M}_{dd}(s)$. This reasoning essentially assumes that
the disk instantaneously adjusts to any changes  of the embryo's mass.

The picture turns out to be 
quite different in reality because the disk cannot quickly 
adapt to the embryo's evolution --- on the contrary it is the 
embryo that initially evolves much faster than the disk
when  ${\cal M}> {\cal M}_{dd}$. As a result,
during the time needed for the embryo's gravity to clear a gap in the disk
its mass grows substantially. Although accretion 
in this part of $s$-${\cal M}$ space is not in the runaway 
regime it is still quite fast initially; only after the embryo gains 
a considerable amount of mass the accretion 
smoothly changes its character to an orderly growth with 
$T_M\approx T_{em}$ in the region between the curves 
${\cal M}= {\cal M}_{dd}$ and ${\cal M}={\cal M}_{M-em}$. Thus, accretion
sets into an orderly mode not when $T_{em}\approx T_{pl}$ 
(as analysis with a fixed embryo mass would assume) 
but only when $T_{em}\approx T_{M}$ which implies larger 
critical mass of the embryo\footnote{As we have mentioned above this 
increase is not big enough to resolve the timescale issue.}.

Strong increase of the planetesimal velocity in the heated zone compared
to that at infinity can be an important issue
for the planetesimal growth in this region. Indeed, planetesimal coagulation 
in this part of the disk would almost certainly 
proceed in the orderly regime
(in contrast to the embryo's accretion): planetesimal velocities 
should be so high that
the gravitational focussing is negligible
and accretion cross-section is independent of planetesimal velocities 
[see equation (\ref{eq:orderly})]. Planetesimal agglomeration
in the excited zone would then be so slow that the growth of more massive 
bodies in the heated region might essentially be chocked (planetesimal 
fragmentation would only strengthen this conclusion). 
New planetary embryos would
be able to emerge only far enough from the preexisting ones 
where the planetesimal dynamics is not strongly affected by 
the gravitational perturbations of massive bodies.

It is also interesting to note that gravitational focussing is always
important for the embryo's accretion (unlike the case of 
orderly growth, see \S \ref{sect:intro4}). Focussing is weak 
when the planetesimal velocities are larger 
than the escape speed from the embryo's surface or when 
$S>p^{-1/2}\gg 1$ or $s>{\cal M}^{1/3}p^{-1/2}$
[see (\ref{eq:accr_rate_num})]. In Figures 
\ref{fig:evol_2.6_2.4}-\ref{fig:evol_3_4} we plot this restriction with a 
long-dash-dotted line ($S=p^{-1/2}$). 
One can see that evolutionary tracks never 
penetrate into this region: although initially $S$ increases towards 
$p^{-1/2}$ as planetesimal velocity grows, the rapid increase of the 
embryo's mass caused by the runaway accretion makes $S$ drop significantly
(sometimes bringing embryo-planetesimal interaction into the
shear-dominated regime) 
when the condition $T_{M}<T_{pl}$ is fulfilled, see 
\S \ref{subsect:regime_separation_all} and equation (\ref{eq:m_el_eredict}).
This implies that the embryo's 
accretion cross-section is {\it always} strongly increased over its geometric 
value by the gravitational focussing\footnote{This also means that embryo's
erosion by planetesimals colliding with it is unimportant.}. 
An important conclusion which 
can be drawn from this is that the growth of the embryo  
{\it never} proceeds in the
orderly fashion with weak gravitational focussing,
as proposed by Safronov (1972). 
Instead, it starts as a runaway evolution 
(Wetherill \& Stewart 1989) which then switches to an
oligarchic growth similar to that predicted by Kokubo \& Ida (1998).


\subsection{Limitations of the analysis.}
\label{subsect:limitations}


The importance of our conclusions relies on the applicability of our
simple model to the realistic protoplanetary disks. There is a lot of
complications which can potentially affect our results. The assumption of
a single mass population is certainly one of them. 
Our whole analysis relies on the possibility to compress the details
of the planetesimal mass spectrum into properties of 
a single population with a unique 
characteristic mass. 
This is only possible if the estimates of such 
averaged mass characterizing separately the embryo's 
accretion and disk dynamical evolution agree with 
each other. If the distribution of masses and random velocities is such
that this condition is not fulfilled
(i.e. if embryo's accretion and disk dynamics are dominated by different 
parts of the planetesimal mass spectrum) then our consideration would fail
and one would need to resort to a more intricate analysis.

Inclusion of a spectrum of planetesimal masses  would 
also allow
dynamical friction to redistribute the random energy of epicyclic motion
between planetesimal populations with 
different masses (Stewart \& Wetherill 1988).
This might lead to an interesting situation when the embryo's interaction 
with different populations proceeds in different dynamical regimes ---
some will be in the shear-dominated regime while some 
in the dispersion-dominated.  
This can introduce some quantitative corrections to our results but we
would expect the general picture of the planetary growth outlined here to 
remain unchanged: we have demonstrated in 
\S \ref{subsect:regime_separation_all}
that even with a single mass population of planetesimals we 
can still adequately reproduce the major features of 
the embryo's runaway growth. 

Another potential worry is the {\it evolution} of the mass spectrum.
This leads to the evolution of the average planetesimal mass $m$
and affects $N a r_H$ as a result. Note that changes of $m$ can be 
caused not only by planetesimal coagulation which always tends to 
increase $m$ but also by the change of the {\it shape} of the planetesimal
mass spectrum driven by the embryo's presence. The last 
possibility might
in principle lead to a {\it decrease} of $m$ if the control over 
the embryo-disk evolution switches to a lower-mass part of the 
planetesimal spectrum. Evolution of $m$ is
probably  
not very important initially because in the runaway regime the 
embryo grows faster than planetesimals, see equation (\ref{eq:runaway}). 
But later on when embryo dominates the disk dynamics and the system 
settles into the asymptotic regime with very slow accretion the 
planetesimal mass evolution might lead to some new effects. 
In addition, the growth of $m$ is tightly coupled to the damping 
of the planetesimal epicyclic motion by inelastic scattering because 
both damping and coagulation are consequences of the same process ---
direct planetesimal collisions. Preliminary calculations demonstrate 
however that
the decrease of planetesimal velocities by inelastic collisions would not 
allow embryo's accretion to continue in the runaway regime until the 
isolation mass is reached  --- at some point it 
would still switch to an orderly regime where ${\cal M}$ grows 
as some power law of
time,  although this orderly growth will be somewhat 
faster than that predicted 
by our analysis of \S \ref{subsect:regime_separation_all} 
[see (\ref{eq:sols_em})]. Planetesimal fragmentation can also become important
when relative planetesimal velocities exceed the escape speed from 
the planetesimal surface (roughly when $s\ge p^{-1/2}$).
One can extend our analysis and numerical 
calculations to include at least the most important of these effects  
into account and we are going to do this in 
the future. 

Our consideration has also been restricted to studying
only a single embryo in the 
disk. In reality there would probably be many of them, at least initially.
Since their masses grow by several orders of magnitude the Hill radii 
of neighboring embryos (as well as their 
excited regions) would overlap at some point. This
would homogenize the disk to some extent but it would 
also slow down the accretion even more because 
disk would be heated by 
several embryos at a time. Finally gravitational perturbations 
between embryos would lead to their collisions and agglomeration. 
This question  certainly deserves more attention both from the 
theoretical and from the numerical points of view. 
One should also be concerned with 
the possibility of embryo's migration which can in principle replenish the
feeding zone of the embryo (Tanaka \& Ida 1999; Ida \etal 2000).

Finally, gas drag can be important for the planetesimal dynamics 
especially for small planetesimal masses (Adachi \etal 1976). 
Its effect is similar to 
that of inelastic collisions --- it damps epicyclic 
motions of planetesimals and leads to somewhat faster accretion  by the 
embryo in the asymptotic regime. 
Gas drag can naturally be incorporated into our
analytical apparatus of Paper II and its effects 
will be explored later on.


\section{Conclusions.}
\label{sect:conclusion}


In this paper we have self-consistently studied the dynamical evolution of
a planetesimal disk coupled to the growth of a single massive 
protoplanetary embryo. We are able to demonstrate that the evolution
of the embryo-disk system proceeds in two consecutive stages:
\begin{itemize}
\item It starts with a rapid runaway growth of the embryo, during which
its mass is not large enough to affect the disk dynamics and the 
spatial distribution
of planetesimals. On a rather short timescale (comparable to that arising 
in conventional coagulation simulations) the embryo reaches the
mass at which it takes over the control of the disk heating and 
runaway growth stops.
\item After a short 
transient stage, the embryo-disk system settles into the asymptotic regime in
which the timescale of the embryo's mass growth is comparable to the 
timescale on which the disk is heated by the embryo's gravity.
\end{itemize}
During the last stage, the embryo dynamically excites planetesimal epicyclic
motions in its vicinity and repels the planetesimals 
forming a depression in
the surface density of the guiding centers (see Figure \ref{fig:time_ev}). 
This effect produces a negative feedback for the embryo's accretion rate
for two reasons: 
rapid growth of the random velocities of planetesimals leads to weaker
gravitational focussing while the decrease of the instantaneous surface 
density (see Figure \ref{fig:ninst_3_3}) reduces the amount of material 
available for accretion. As a result, growth of the protoplanetary embryo 
proceeds slower than in homogeneous planetesimal disks. This implies that
conventional ``particle-in-a-box'' coagulation simulations 
(Wetherill \& Stewart 1989; Kenyon \& Luu 1998; Inaba \etal 2001) might be 
underestimating the timescale of the 
protoplanetary embryo's growth, sometimes
by rather large factors (up to $\sim 10$). Multi-zone coagulation simulations 
(Spaute \etal 1991; Weidenschilling \etal 1997)
should give a more reliable description of the protoplanetary 
disk evolution.

We have presented our equations in a dimensionless form which has
allowed us to 
uncover a very important property of the problem we consider: its outcome
depends on only two dimensionless parameters --- the number of planetesimals
inside the annulus of the disk with width equal to the planetesimal Hill
radius $r_H$, $N a r_H$, and the ratio of the physical to Hill 
radii of a solid body (planetesimal or embryo) $p$. All astrophysically
relevant characteristics of the system ($m$, $a$, $M_c$, $\Sigma$, $\rho$)
are combined into these two parameters which 
greatly simplifies the exploration of the
parameter space. We have numerically studied in \S \ref{sect:numerical} the 
evolution of the system for a number of pairs 
of $N a r_H$ and $p$; combined with our 
analytical developments in \S \ref{sect:scaling} these results allow
us to formulate a set of simple scaling relations 
(\ref{eq:star_eredict})-(\ref{eq:m_tau_eredict})
which can be used to predict the embryo-disk evolution for different 
sets of $N a r_H$ and $p$.

To summarize we can say that the embryo's growth starts in the runaway regime,
but then switches into the  ``oligarchic''
growth of Kokubo \& Ida (1998). We have demonstrated that the 
embryo's accretion never proceeds in the orderly regime with weak
gravitational focussing suggested by Safronov (1972) --- planetesimal
velocities are always smaller than the escape velocity from the embryo's
surface. The embryo formation timescale 
and some details of the embryo-disk evolution are 
subject to a number of uncertainties related
to the simplicity of our model, as we discussed in \S \ref{sect:discussion}.
Nevertheless, we believe that our major conclusions are robust:
the results of our qualitative analysis of \S \ref{sect:scaling}
and numerical calculations presented in \S \ref{sect:numerical} 
suggest that the dynamical interaction between the 
protoplanetary embryo and planetesimal disk is a very important issue 
which should certainly be addressed when realistic 
coagulation simulations are performed. 
The general agreement between the results of our simple  analysis
of \S \ref{sect:scaling} with more accurate numerical calculations of 
\S \ref{sect:numerical} encourages the studies of more complex and more 
realistic problems in the same spirit. In the future we are going to 
improve our treatment of the coupled planetesimal disk and embryo evolution 
by relaxing simplifying assumptions employed in this paper. We are also 
going to include additional mechanisms which are important
in modelling of planetary formation --- effects of multiple embryos,
migration, gas drag, etc. 

\acknowledgements

I am grateful to Scott Tremaine for his advice, and for 
reading this manuscript 
and making a lot of useful suggestions. The 
financial support of this work by 
 the Charlotte Elizabeth Procter Fellowship and NASA grant NAG5-10456
is thankfully acknowledged.

\appendix


\section{More accurate equations of evolution.}  
\label{sect:exact}


Here we describe the system of equations we are going to use 
to follow  the 
evolution of planetesimal surface density, eccentricity and inclinations. 
We derive these equations for
the general case of a distribution of planetesimal masses for future 
applications. For this reason here $N$ is a function of planetesimal 
mass\footnote{Unlike \S \ref{sect:scaling} where we have studied
only a single mass population. We go back to this simpler case later on.} $m$.

We will use the following notation. First of all, we introduce
the dimensionless radial distance from the embryo's orbit 
$H$, normalized by its Hill radius:
\begin{eqnarray}
H=\Delta a/R_H,~~~R_H=a\mu_e^{1/3},~~~\mu_e=M_e/M_c,
\label{eq:H_def}
\end{eqnarray}
where $\Delta a$ is the dimensional radial distance from the
 embryo's orbit (equivalent to $h a$ in the notation
of Paper II).
We also scale all eccentricities and inclinations by $R_H/a=
\mu_e^{1/3}$. This gives us new variables $S$ and $S_z$ defined
by equation (\ref{eq:new_vels}).
When doing this we have to keep in mind that $S$ and $S_z$ can vary
not only because eccentricity and inclinations 
$\sigma_e$ and $\sigma_i$ vary with time but also because
$\mu_e$ can change  with time (as a consequence of planetary mass growth).
This variation can be easily taken into account by noticing that
\begin{eqnarray}
\frac{\partial}{\partial t}\left(N\{S,S_z\}^2\right)=
\frac{1}{\mu_e^{2/3}}\frac{\partial}{\partial t}
\left(N\sigma_{\{e,i\}}^2\right)-
N\sigma_{\{e,i\}}^2~\frac{2}{3}\frac{\partial \ln\mu_e}{\partial t}.
\label{eq:switch}
\end{eqnarray}

We will characterize mass distribution of planetesimals by some
fiducial mass $m_\star$ and introduce a dimensionless mass $z$ such that
\begin{eqnarray}
z=m/m_\star=\mu/\mu_\star,~~~\mu=m/M_c,~\mu_\star=m_\star/M_c.
\label{eq:z_def}
\end{eqnarray}
Then ${\cal M}=M_e/m_\star$
Associated with this mass is a fiducial surface number 
density $N_\star=\Sigma_0/m_\star$, where $\Sigma_0$ is
the total surface {\it mass} density of the disk. Following Papers II
\& III we assume both $N(m)$ and $\Sigma_0$ to be scaled by $a^2$
(so that $N(m)=N(z)$ is dimensionless, unlike \S \ref{sect:scaling}). 
Finally we define a new surface number density function
\begin{eqnarray}
g(z,H)=m_\star\frac{N(z)}{N_\star}=m_\star^2\frac{N(z)}{\Sigma_0},
~~~~~~~~~~\int\limits_0^\infty g(z,H) z dz=1.
\label{eq:eq:g_def}
\end{eqnarray}
We will use the notation $g_i(H)$ for the surface 
number density of planetesimals
of mass $m_i$ (or $z_i=m_i/m_\star$).

In transforming the planetesimal-planetesimal part of the evolution
equations we will take into account that scattering coefficients in
the dispersion-dominated regime have a rather specific form given by equations 
(82)-(93) of Paper II; we need to remember that expressions 
for these coefficients 
are functions of $\sigma_{e,i}$
normalized by Hill radius of planetesimal-planetesimal interaction $r_H$,
which is different from $S$ and $S_z$.
With this in mind, using definitions (\ref{eq:H_def})-(\ref{eq:eq:g_def})
and equations (49)-(55) of Paper II and (13)-(15), (24)-(26), (B2) 
of Paper III
we can finally represent equation of evolution of 
quantity $F=\{g,gs^2,gs_z^2\}$ in the following general form:
\begin{eqnarray}
\frac{\partial F}{\partial \tau}=\pi
N_\star\mu_\star^{1/3}{\cal M}^{-4/3}
\frac{\partial F}{\partial \tau}\bigg|_{pl}+
{\cal M}^{1/3}\left[
\theta_1\frac{\partial F}{\partial \tau}\bigg|_{hot}+
\theta_2\frac{\partial F}{\partial \tau}\bigg|_{cold}\right]-
\frac{2}{3}F\frac{\partial \ln{\cal M}}{\partial \tau}
\label{eq:most_general}
\end{eqnarray}
with a new dimensionless time $\tau=(3/4\pi)\Omega t\mu_\star^{1/3}=t/t_{syn}$
[see (\ref{eq:change_of_var})]. 
The interpolating functions $\theta_1=\Theta(S,S_z)$ and 
$\theta_2=1-\Theta(S,S_z)$ provide smooth matching between the shear-
and dispersion-dominated regimes of embryo-planetesimal interaction
(marked with subscripts ``cold'' and ``hot'', see Paper III). 
Function $\Theta(S,S_z)$ has the following properties:
$\Theta(S,S_z)\to 1$ as $S, S_z\to \infty$, and 
$\Theta(S,S_z)\to 0$ as $S, S_z\to 0$. 
The last term in (\ref{eq:most_general}) is present only in equations 
for eccentricity and inclination evolution.

For the surface density of planetesimals of type 1, $F=g_1$ and we have 
\begin{eqnarray}
&& \frac{\partial g_1}{\partial \tau}\bigg|_{pl}=
-\frac{\partial}{\partial H}\left(\Upsilon^N_1 g_1\right)
+\frac{\partial^2}{\partial H^2}\left(\Upsilon^N_2 g_1\right),
\label{eq:pl_el_surf}\\
&& \frac{\partial g_1}{\partial \tau}\bigg|_{hot}=
-\frac{\partial}{\partial H}\left(|H|\langle\Delta\tilde h\rangle
g_1\right)+
\frac{\partial^2}{\partial H^2}\left(|H|\langle(\Delta\tilde h)^2\rangle g_1
\right),
\label{eq:em_surf}\\
&& \frac{\partial g_1}{\partial \tau}\bigg|_{cold}=
-g_1(H)|H|+
\int\limits_{-\infty}^{\infty}
\tilde P(H_0,H)g_1(H_0)|H_0|dH_0,
\label{eq:surf_shear}
\end{eqnarray}
where
\begin{eqnarray}
&& \Upsilon^N_1(H)=
2\int\limits_0^{\infty}dz_2 z_2(z_1+z_2)
\int\limits_{-\infty}^{\infty}d H_1 |H_1|
\langle\Delta \tilde h\rangle
g_2(H-H_1),
\label{eq:Ups_1_surf}\\
&& \Upsilon^N_2(H)=
\int\limits_0^{\infty}dz_2 z_2^2
\int\limits_{-\infty}^{\infty}d H_1 |H_1|
\langle(\Delta \tilde h)^2\rangle
g_2(H-H_1),
\label{eq:Ups_2_surf}
\end{eqnarray}
where $z_2$ is used as a variable of integration
over the planetesimal mass spectrum. 

For eccentricity evolution, $F=2g_1 S_1^2$ and  
\begin{eqnarray}
&& \frac{\partial}{\partial \tau}\left(2g_1 S_1^2\right)\bigg|_{pl}=
\Upsilon^e_0 g_1-\frac{\partial}{\partial H}\left(\Upsilon^e_1 g_1\right)
+\frac{\partial^2}{\partial H^2}\left(\Upsilon^e_2 g_1\right),
\label{eq:pl_el_ecc}\\
&& \frac{\partial}{\partial \tau}\left(2g_1 S_1^2\right)\bigg|_{hot}=
|H|\langle\Delta ({\bf \tilde e}^2)\rangle g_1
-\frac{\partial}{\partial H}\left(|H| 
\langle({\bf \tilde e}^2+2{\bf \tilde e}
\cdot\Delta {\bf \tilde e})\Delta \tilde h\rangle g_1\right)\nonumber\\
&& +\frac{\partial^2}{\partial H^2}\left(|H| 
\langle{\bf \tilde e}^2(\Delta \tilde h)^2\rangle g_1\right),
\label{eq:em_s_e}\\
&& \frac{\partial }{\partial \tau}\left(2g_1 S_1^2\right)\bigg|_{cold}=
-2g_1(H)S_1^2(H)|H|
\nonumber\\
&& +\int\limits_{-\infty}^{\infty}
\tilde P(H_0,H)g_1(H_0)\left(2s_1^2(H_0)+\frac{3}{4}
\Delta (\tilde {\bf e}_{sc})^2(H_0)\right)|H_0|dH_0,
\label{eq:ecc_shear}
\end{eqnarray}
where
\begin{eqnarray}
&& \Upsilon^e_0(H)=
2\int\limits_0^{\infty}dz_2z_2(z_1+z_2)
\int\limits_{-\infty}^{\infty}d H_1 |H_1|
g_2(H-H_1)\nonumber\\
&& \times\left[\frac{z_2}{z_1+z_2}\langle(\Delta {\bf \tilde e}
)^2\rangle
+2\frac{S_1^2}{S_1^2+ S_2^2}
\langle{\bf \tilde e}\cdot\Delta {\bf \tilde e}\rangle\right],
\label{eq:Ups_0_e}\\
&& \Upsilon^e_1(H)=
2\int\limits_0^{\infty}dz_2
z_2(z_1+z_2)
\int\limits_{-\infty}^{\infty}d H_1 |H_1|
g_2(H-H_1)\nonumber\\
&& \times\left[
2\frac{S_1^2 S_2^2}
{S_1^2+S_2^2}\langle
\Delta \tilde h\rangle+
\frac{S_1^4}
{(S_1^2+S_2^2)^2}
\langle {\bf \tilde e}^2\Delta \tilde h\rangle
+2\frac{z_2}{z_1+z_2}\frac{S_1^2}{S_1^2+ S_2^2}
\langle({\bf \tilde e}\cdot\Delta {\bf \tilde e})\Delta 
\tilde h\rangle\right],
\label{eq:Ups_1_e}\\
&& \Upsilon^e_2(H)=
\int\limits_0^{\infty}dz_2 z_2^2
\int\limits_{-\infty}^{\infty}d H_1 |H_1|
g_2(H-H_1)\nonumber\\
&& \times\left[
2\frac{S_1^2 S_2^2}
{S_1^2+S_2^2}\langle
(\Delta \tilde h)^2\rangle+
\frac{S_1^4}
{(S_1^2+S_2^2)^2}
\langle {\bf \tilde e}^2(\Delta \tilde h)^2\rangle\right],
\label{eq:Ups_2_e}\\
&& S_1=S_1(H),~~~S_2=S_2(H-H_1).
\end{eqnarray}

Different terms in the equation of evolution of
$2g_1 S_{z 1}^2$ look analogous to (\ref{eq:pl_el_ecc})-(\ref{eq:em_s_e})
and (\ref{eq:Ups_0_e})-(\ref{eq:Ups_2_e}) if we replace $S$ with $S_z$
but the shear-dominated part reads (see Paper III)
\begin{eqnarray}
\frac{\partial}{\partial \tau}\left(2g_1S_{z 1}^2\right)\bigg|_{cold}=
-2g_1(H)S_{z 1}^2(H)|H|-2
\int\limits_{-\infty}^{\infty}
\tilde P(H_0,H)g_1(H_0)S_{z 1}^2(H_0)
|H_0|dH_0.
\label{eq:inc_shear}
\end{eqnarray}

In (\ref{eq:surf_shear}), (\ref{eq:ecc_shear}), (\ref{eq:inc_shear}) 
$H_0$ is an integration variable having the meaning of 
initial difference in semimajor axes of interacting bodies and 
\begin{eqnarray}
\tilde P(H_0,H)=\delta\left[H-H_0-\Delta 
\tilde h_{sc}(H_0)\right],
\label{eq:probab}
\end{eqnarray}
where $\tilde h_{sc}(H_0)$ is some function for which the analytical 
prescription based on the results of numerical calculations exists
(see Petit \& H\'enon 1987).
Scattering coefficients 
$\langle\Delta\tilde h\rangle$,
$\langle(\Delta\tilde h)^2\rangle$,
$\langle{\bf \tilde e}^2\Delta \tilde h\rangle$,
$\langle({\bf \tilde e}\cdot
\Delta {\bf \tilde e})\Delta \tilde h\rangle$,
$\langle{\bf \tilde e}^2(\Delta \tilde h)^2\rangle$,
$\langle(\Delta {\bf \tilde e})^2\rangle$,
$\langle{\bf \tilde e}\cdot\Delta {\bf \tilde e}\rangle$
in these equations are functions
of $S,S_z$, e.g.
\begin{eqnarray}
&&\langle \Delta \tilde h\rangle=-\frac{4}{3}\frac{\ln\Lambda}
{S_r^2 S_{z r}^2}e^{-H^2/(2 S_r^2)}
\frac{1}{H}U_0^0\left(\frac{H/S_r}{2\sqrt{2}}
\frac{H/S_{z r}}{2\sqrt{2}}\right),\nonumber\\
&&\Lambda={\cal M}^{1/3}\frac{S_{z r}}{z_1+z_2}
(c_1 S_r^2 + c_2 S_{z r}^2)\gg 1,~~~~~
S_r=\sqrt{S_1^2+S_2^2},~
S_{z r}=\sqrt{S_{z 1}^2+S_{z 2}^2},
\label{eq:av_dh}
\end{eqnarray}
where $c_1$ and $c_2$ are some constants (their numerical values are 
fixed approximately in Paper III), and function $U_\rho^{\eta}$
is defined by equation (95) of Paper II. Scattering coefficients describing
embryo-planetesimal interaction are analogous to (\ref{eq:av_dh}) but
have  $\Lambda=(c_1 S_r^2 + c_2 S_{z r}^2)S_{z r}/(z_1+z_2)$.

For the growth of the planetary embryo's mass we can write using 
equations (42)-(44) of Paper III:
\begin{eqnarray}
&& \frac{\partial {\cal  M}}{\partial \tau}=
\frac{4}{3}p N_\star \mu_\star^{1/3}{\cal M}^{2/3}
\int\limits_0^\infty zdz\nonumber\\
&& \times
\left[\frac{\pi_1}{8}
\int\limits_{-\infty}^\infty g(z,H)\frac{|H|}{S^2 S_z^2}e^{-H^2/(2 S^2)}
\left(\frac{p}{4}H^2 U_+ +2 U_-\right)dH
+\frac{5\pi_2}{S_z} g^{inst}(z)\right],
\label{eq:accr_rate_num}
\end{eqnarray}
where functions $\pi_1=\Pi(S,S_z)$ and 
$\pi_2=1-\Pi(S,S_z)$ are used to interpolate
between the shear-
and dispersion-dominated regimes of embryo-planetesimal interaction. 
Function $\Pi(S,S_z)$ has properties analogous to the properties of 
$\Theta(S,S_z)$ (see Paper III). 
The quantity $g^{inst}$ in the second term in brackets is the instantaneous
surface density of planetesimals at the embryo's location, which
can be calculated using the conversion between $N$ and $N^{inst}$ described
in Paper III. The dimensionless parameter $p$ was defined in 
(\ref{eq:change_of_var}).
The first terms in the square bracket describe the mass accretion rate in the 
dispersion-dominated and 
shear-dominated regime, respectively.

When a single mass population of planetesimals is considered one should
use the planetesimal mass $m$ as a fiducial mass $m_\star$ and set 
$g(z,H)=\delta(z-1)f(H)$, where $f(H)$ is a function of distance $H$
only. Then the integration over $z_2$ in (\ref{eq:Ups_1_surf}),
(\ref{eq:Ups_2_surf}), (\ref{eq:Ups_0_e})-(\ref{eq:Ups_2_e})
trivially goes away. In this work we have only considered a 
single mass population of planetesimals.

One can easily notice that in agreement with the results of 
\S \ref{sect:scaling}
there are only two free parameters present in equations
(\ref{eq:most_general})-(\ref{eq:accr_rate_num}): $p$ --- ratio 
of physical radius to the Hill radius of any object,
and $N_\star \mu_\star^{1/3}$ --- measure of the disk surface number density
equivalent to $N a r_H$ of \S \ref{sect:scaling} (in 
\S \ref{sect:scaling} $N$ is dimensional).
When the distribution of planetesimal masses exists the shape of
the mass spectrum is also a prescribed input function (which can 
vary in time as a result of coagulation).
Equations (\ref{eq:most_general})-(\ref{eq:accr_rate_num}) constitute a closed
set of equations describing the behavior of the embryo-disk system. Their
 numerical solutions are presented in \S \ref{sect:numerical}.

\end{document}